%% file: BruknerZukowski-BellInequalitiesandQuantumCommunication.tex
\documentclass[graybox]{svmult}

\usepackage{mathptmx}       
\usepackage{helvet}         
\usepackage{courier}        
\usepackage{type1cm}        
\usepackage{makeidx}         
\usepackage{graphicx}        
\usepackage{multicol}        
\usepackage[bottom]{footmisc}

\makeindex             

\newcommand{\ket}[1]{\left | \, #1 \right \rangle}
\newcommand{\BEQ}{\begin{equation}}
\newcommand{\EEQ}{\end{equation}}
\newcommand{\BEQA}{\begin{eqnarray}}
\newcommand{\EEQA}{\end{eqnarray}}

\begin{document}

\title*{Bell's Inequalities: Foundations and Quantum Communication}
\author{{\v C}aslav Brukner and Marek \.Zukowski}
\institute{{\v C}aslav Brukner \at Faculty of
Physics, University of Vienna, Boltzmanngasse 5, 1090 Vienna; \\
Institute of Quantum Optics and Quantum Information, Austrian
Academy of Sciences, Boltzmanngasse 3, 1090 Vienna\\
\email{caslav.brukner@univie.ac.at} \and Marek \. Zukowski \at
Institute for Theoretical Physics and Astrophysics, University of
Gdansk, 80-952 Gdansk, Poland \email{marek.zukowski@univie.ac.at}}
%
%
\maketitle

\abstract{ For individual events quantum mechanics makes only
probabilistic predictions. Can one go beyond quantum mechanics in
this respect? This question has been a subject of debate and
research since the early days of the theory. Efforts to construct
deeper, realistic, level of physical description, in which
individual systems have, like in classical physics, preexisting
properties revealed by measurements are known as hidden-variable
programs. Demonstrations that a hidden-variable program necessarily
requires outcomes of certain experiments to disagree with the
predictions of quantum theory are called ``no-go theorems''. The
Bell theorem excludes local hidden variable theories. The
Kochen-Specker theorem excludes noncontextual hidden variable
theories. In local hidden-variable theories
faster-that-light-influences are forbidden, thus the results for a
given measurement (actual, or just potentially possible) are
independent of the settings of other measurement devices which are
at space-like separation. In noncontextual hidden-variable theories
the predetermined results of a (degenerate) observable are
independent of any other observables that are measured jointly with
it.}

It is a fundamental doctrine of quantum information science that
quantum communication and quantum computation outperforms their
classical counterparts. If this is to be true, some fundamental
quantum characteristics must be behind better-than-classical
performance of information processing tasks. This chapter aims at
establishing connections between certain quantum information
protocols and foundational issues in quantum theory. After a brief
discusion of the most common misinterpretations of Bell's theorem
and a discussion of what its real me aning is, it will be
demonstrated {\it how quantum contextuality and violations of local
realism can be used as useful resources} in quantum information
applications. In any case, the readers should bear in mind that this
chapter is not a review of the literature of the subject, but rather
a quick introduction.

\section{Introduction}

Which quantum states are useful for quantum information processing?
All non-separable states? Only distillable non-separable states?
Only those which violate constraints imposed by local realism?
Entanglement is the most distinct feature of quantum physics with
respect to the classical world~\cite{schroedinger}. On one hand,
entangled states violate Bell inequalities, and thus rule out local
realistic explanation of quantum mechanics. On the other hand, they
enable certain communication and computation tasks to have an
efficiency not achievable by the laws of classical physics.
Intuition suggests that these two aspects, the fundamental one, and
the one associated with applications, are intimately linked. It is
natural to assume that the quantum states which allow the no-go
theorems of quantum theory, such as Kochen-Specker, Bell's or
Greenberger-Horne-Zeilinger theorem should also be useful for
quantum information processing. If this were not true, one might
expect that the efficiency of quantum information protocols could be
simulatable by classical, essentially local realistic or
noncontextual models, and thus achievable already via classical
means. This intuitive reasoning is supported by the results of, for
example, Acin {\it et. al}~\cite{acin}: violation of a Bell's
inequality is a criterium for the security of quantum key
distribution protocols. Also it was shown that violation of Bell's
inequalities by a quantum state implies that pure-state entanglement
can be distilled from it~\cite{distill} and  that Bell's
inequalities are related to optimal solutions of quantum state
targeting~\cite{bechmann}. In this overview we will give other
examples that demonstrate the strong link between fundamental
features of quantum states and their applicabilities in quantum
information protocols, such as in quantum communication complexity
problems, quantum random access, or certain quantum games.

\section{Quantum predictions for two qubits systems}

To set the stage for our story let us first describe two-qubits
systems in full detail.

We shall present predictions for all possible local yes-no
experiments on two spin-1/2 systems(in modern terminology, qubits)
for all possible quantum states, i.e. from the pure maximally
entangled singlet state (or the Bohm-EPR state), via factorizable
(i.e. non-entangled) states, up to any mixed state. This will enable
us to reveal the distinguishing traits of the quantum predictions
for entangled states of the simplest possible compound quantum
system. The formalism can be applied to any system consisting of two
subsystems, such that each of them is described by a two dimensional
Hilbert space. We choose the spin-$1/2$ convention to simplify the
description.

\subsection{Pure states}

An important tool simplifying the analysis of the pure states of two
subsystems is the so-called Schmidt decomposition.

\subsubsection{Schmidt decomposition}

For any nonfactorizable (i.e., entangled) pure state, $|\psi\rangle$
of {\it pair} of quantum subsystems, one described by a Hilbert
space of dimension $N$, the other by space of dimension $M$, $N\leq
M$, it is always possible to find preferred bases, one basis for the
first system, another one for the second, such that the state
becomes a sum of bi-orthogonal terms, i.e. \BEQ
\ket{\psi}=\sum_{i=1}^Nc_i\ket{a_i}_1\ket{b_i}_2 \label{Schmidt}
\EEQ with ${}_n\langle x_i|x_j\rangle_n=\delta_{ij}$, for $x=a,b$
and $n=1,2$. It is important to stress that the appropriate single
subsystem bases, here $\ket{a_i}_1$ and $\ket{b_j}_2$, depend upon
the state that we want to Schmidt-decompose.

The ability to Schmidt decompose the state is equivalent to a well
known fact form matrix algebra, that any $N\times M$ matrix ${\bf
{\hat A}}$ can be always put into a diagonal form ${\bf \hat D}$, by
applying a pair of unitary transformations:
$\sum_{j=1}^{N}\sum_{k=1}^MU_{ij}A_{jk}U_{kl}=D_l\delta_{il}$.

The interpretation of the above formula could be put as follows. If
the quantum pure state of two systems is non-factorizable, then
there exist a pair of local observables (for system $1$ with
eigenstates $|a_i\rangle$, and for system $2$ with eigenstates
$|b_i\rangle$) such that the results of their measurement are
perfectly correlated.

The method of Schmidt decomposition allows one to put every pure
normalized state of two spins into \BEQ
|\psi\rangle=\cos{\alpha/2}\ket{+}_1\ket{+}_2
+\sin{\alpha/2}\ket{-}_1\ket{-}_2 \label{schmidtcoeff}. \EEQ Schmidt
decomposition generally allows the coefficients to be real. This is
achievable via trivial phase transformations of the preferred bases.

\subsection{Arbitrary states}

Systems can be in mixed states. Such states describe situations in
which there does not exist any {\em nondegenerate} observable for
which measurement result is deterministic. This is the case when the
system can be with various probabilities $P(x)\geq0$ in some
non-equivalent states $|\psi(x)\rangle$, with $\sum_xP(x)=1$. Mixed
states are represented by self adjoint density non-negative
operators $\varrho = \sum_xP(x)|\psi(x)\rangle \langle \psi(x)|$. As
${\rm Tr}|\psi(x)\rangle\langle\psi(x)|=1$ one has ${\rm
Tr}\varrho=1$.

Let us present in detail properties of mixed states of the two
spin-$1/2$ systems. Any self adjoint operator for one spin-$1/2$
particle is a linear combination of the Pauli matrices $\sigma_{i}$,
$i=1,2,3$ and the identity operator, $\sigma_{0}=\mathbf{1}$, with
{\it real} coefficients. Thus, any self adjoint operator in the
tensor product of the two spin-$1/2$ Hilbert spaces, must be a real
linear combination of all possible products of the operators
$\sigma_{\mu}^1\sigma_{\nu}^2$, where the Greek indices run from $0
$ to $3$, and the superscripts denote the particle.  As the trace of
$\sigma_i$  is zero we arrive at the following form of  the general
density operator for two spin $1/2$ systems:
\begin{equation}
\varrho={1\over4}\left(\sigma^{(1)}_0 \sigma^{(2)}_0 + \bf{
r\cdot\sigma^{(1)}}\sigma^{(2)}_0 +\sigma^{(1)}_0 \bf{s\cdot\sigma^{(2)}}\right. +
\left.\sum_{m,n=1}^3T_{nm}\sigma_{n}^{(1)}\sigma_{m}^{(2)}\right),
\label{postac1}
\end{equation}
where, $ {\bf r}$, ${\bf s}$ are real three dimensional vectors and
${\bf r}\cdot\sigma\equiv\sum_{i=1}^3 r_{i}\sigma_{i}$. We shall use
the tensor product symbol $\otimes$ only sparingly, only whenever it
is deemed necessary.  The condition ${\rm Tr}\varrho=1$ is satisfied
thanks to the first term.

Since the average of any real variable which can have only two
values $+1$ and $-1$ cannot be larger than $1$ and less than $-1$,
the real coefficients $T_{mn}$ satisfy relations \BEQ -1\leq
T_{mn}={\rm Tr}\varrho\sigma_n^{(1)} \sigma_m^{(2)}\leq1,
\label{defT} \EEQ and they form a  matrix which will be denoted by
$\bf{\hat{T}}$. One also has \BEQ -1\leq r_n={\rm
Tr}\varrho\sigma_n^{(1)} \leq1, \label{defr} \EEQ and \BEQ -1\leq
s_{m}={\rm Tr}\varrho \sigma_m^{(2)}\leq1. \label{defs} \EEQ

\subsubsection{Reduced density matrices for subsystems}

A reduced density matrix represents the local state of a compound
system. If we have two subsystems, then the average of any
observable which pertains to the first system only, i.e. of the form
$A\otimes \mathbf{1}$, where $\mathbf{1}$ is the identity operation
for system $2$, can be expressed as follows ${\rm Tr}_{12}(A\otimes
\mathbf{1} \varrho)={\rm Tr}_1[A({\rm Tr}_2\varrho)].$ Here ${\rm
Tr}_i$ represents a trace with respect to system $i$. As trace is a
basis independent notion, one can always choose a factorizable
basis, and therefore split the trace calculation into two stages.

The reduced one particle matrices for spins $1/2$, are of the
following form:
\begin{eqnarray}
\varrho_1\equiv{\rm {\rm Tr}}_{2}\varrho={1\over2}(\mathbf{1} +\bf{
r\cdot\sigma^{(1)}}),\\
\varrho_2\equiv{\rm {\rm Tr}}_{1}\varrho={1\over2}(\mathbf{1}+
\bf{s\cdot\sigma^{(2)}}). \label{redukcje}
\end{eqnarray}
with $\vec{r}$ and $\vec{s}$ the two local Bloch vectors of the
spins.

Let us denote the eigenvectors  of the spin projection along
direction $\bf{a}$ of the first spin as: $\ket{\psi(\pm1,{\bf
a})}_{1}$. They are defined by the relation \BEQ
\bf{a\cdot\sigma^{(1)}}\ket{\psi(\pm1,{\bf
a})}_1=\pm1\ket{\psi(\pm1,{\bf a})}_1,\EEQ where $\bf {a}$ is a real
vector of unit length (i.e. $\bf{a\cdot\sigma^1}$ is a Pauli
operator in the direction of $\bf{a}$). The  probability of a
measurement of this Pauli observable to give a result $\pm 1$ is
given by \BEQ P(\pm 1|{\bf{a}})_1={\rm Tr}_1\varrho_1\pi_{\bf{(a,
\pm1)}}^{(1)}=\frac{1}{2}(1\pm\bf{a\cdot r}), \EEQ and it is
positive for arbitrary $\bf{a}$, if and only if, the norm of
$\bf{r}$ satisfies \BEQ |\bf{r}|\leq 1. \EEQ Here $ \pi_{\bf{(a,
\pm1)}}^{(1)}$ is the projector $\ket{\psi(\pm1,\bf{a})}_{11}
\langle\psi(\pm1,{\bf a})|$.

\subsection{Local measurements on two spins}

The probabilities for local measurements to give the result $l=\pm1$
for particle 1 and the result $m=\pm1$ for particle 2, under
specified local settings, ${\bf{a}}$ and ${\bf{b}}$ respectively, of
the quantization axes are given by:
\begin{equation} P(l,m|{\bf{a}},{\bf{b}})_{1,2} = {\rm
Tr}\varrho\pi_{(\vec{a}, l)}^{(1)} \pi_{(\vec{b}, m)}^{(2)}
=\frac{1}{4} \left(1+l\vec{a} \cdot \vec{r}+m \vec{b}\cdot \vec{s}+l
m \vec{a} \cdot {\bf \hat{T}\vec{b}}\right),
\end{equation} where $\bf{\hat{T}b}$ denotes the transformation of the
column vector $\bf{b}$ by the matrix $\bf{\hat{T}}$  (we treat here
Euclidean vectors as column matrices).

One can simplify all these relations by performing suitable local
unitary transformations upon each of the subsystems, i.e. via
factorizable unitary operators $U^{(1)}U^{(2)}$. It is well known
that any unitary operation upon a spin $1/2$ is equivalent to a
three dimensional rotation in the space of Bloch vectors. In other
words, for any real vector $\bf{w}$
\begin{equation}
U({\bf \hat{O}}){\bf w\cdot\sigma} U({\bf \hat{O}})^\dagger =({\bf
\hat{O}}\bf{w})\bf{\cdot\sigma}, \label{pawel}
\end{equation}
where ${\bf \hat{O}}$ is the orthogonal matrix of the rotation.
If the density matrix is subjected to such a transformations on
either spins subsystem, i.e. to the $U^1({\bf \hat{O}}_1)U^2({\bf
\hat{O}}_2)$ transformation, the parameters $\bf r, \bf s$  and $\bf
\hat{T}$ transform themselves as follows
\begin{eqnarray}
&{\bf r'}={\bf \hat{O}}_1 {\bf r},&\nonumber \\
&{\bf s'}= {\bf \hat{O}}_2 {\bf s},&\nonumber \\
&{\bf \hat{T}'}= {\bf \hat{O}}_1 {\bf \hat{T}} {\bf
\hat{O}^T}_2.&
\label{TRANSF}\\ \nonumber
\end{eqnarray}
Thus, for an arbitrary state, we can always choose such factorizable
unitary transformation that the corresponding rotations (i.e.
orthogonal transformations) will diagonalize the correlation tensor
(matrix) ${\bf \hat{T}}$. This can be seen as another application of
Schmidt's decomposition, this time in case of second rank tensors.

The physical interpretation of the above is that one can always
choose two (local) systems of coordinates, one for the first
particle, the other for the second particle, in such a way that the
${\bf \hat{T}}$ matrix will be diagonal.

Let us note that one can decompose the two spin density matrix into:
\begin{equation}
\varrho=\varrho_1\otimes \varrho_2+{1\over4}
\sum_{m,n=1}^3C_{nm}\sigma_{n}^1\otimes \sigma_{m}^2, \label{postac1}
\end{equation}
i.e., it is a sum of the product of the two reduced density matrices
and a term $\bf{\hat{C}}=\bf{\hat{T}}-\bf{rs^T}$ which is
responsible for correlation effects.

Any density operator satisfies the inequality $\frac{1}{d}<
\mbox{Tr}\varrho^2\leq 1$, where $d$ is the dimension of the Hilbert
space in which it acts, i. e.  of the system it describes. The value
of $\mbox{Tr}\varrho^2$ is a measure of the purity of the quantum
state. It is equal to $1$ only for single dimensional projectors,
i.e. the pure states. In the studied case one must have \BEQ |{\bf
r}|^2+|{\bf s}|^2 + ||{\bf\hat{T}}||^2\leq3. \label{trace} \EEQ

For pure states, represented by Schmidt decomposition
(\ref{schmidtcoeff}), ${\bf\hat{T}}$ is diagonal with entries
$T_{xx}=-\sin{\alpha}$, $T_{yy}=\sin{\alpha}$ and $T_{zz}= 1$,
whereas ${\bf r}={\bf s}$, and their $z$ component is non-zero:
$s_z=m_z=\cos{\alpha}$. Thus in case of a maximally entangled states
${\bf\hat{T}}$ has only diagonal entries equal to $+1$ and $-1$. In
the case of the singlet state, \BEQ |\psi\rangle=\frac{1}{\sqrt{2}}
\left(\ket{+}_1\ket{-}_2 -\ket{-}_1\ket{+}_2\right), \label{singlet}
\EEQ which can be obtained from eq.~(\ref{schmidtcoeff}), by putting
$\alpha=-\frac{\pi}{2}$ and rotating one of the subsystems such that
$|+\rangle$ and $|-\rangle$ interchange (This is equivalent to a
$180$ degrees rotation with respect to the axis $x$; See above
(\ref{TRANSF})), the diagonal elements of the correlation tensor are
all $-1$.

\section{Einstein-Podolsky-Rosen Experiment}

In their seminal 1935 paper~\cite{epr} entitled {\it ''Can
quantum-mechanical description of physical reality be considered
complete?''} Einstein, Podolsky and Rosen (EPR) consider quantum
systems consisting of two particles such that, while neither
position nor momentum of either particle is well defined, both the
difference of their positions and the sum of their momenta are both
precisely defined. It then follows that measurement of either
position or momentum performed on, say, particle 1 immediately
implies for particle 2 a precise position or momentum respectively
even when the two particles are separated by arbitrary distances
without any actual interaction between them.

We shall present the EPR argumentation for incompleteness of quantum
mechanics in the language of spins $1/2$. This has been done by Bohm
in 1952. A two qubit example of an EPR state is the singlet
state~(\ref{singlet}). Properties of a singlet can be inferred
without mathematical considerations given above. This is a state of
zero total spin. Thus measurements of the same component of the two
spins must always give opposite values - this is simply the
conservation of angular momentum at work. In terms of the language
od Pauli matrices the product of the local results is then always
$-1$. We have (infinitely many) {\em prefect (anti-)correlations}.
We assume that the two spins are very far away, but nevertheless in
the singlet state.

After the translation into the Bohm's example EPR argument runs as
follows. Here are their premises:
\begin{enumerate}
\item {\it Perfect correlations} If whatever spin components of particles 1 and 2,
then with certainly the outcomes will be found to be perfectly
anti-correlated.
\item {\it Locality}: ''Since at the time of measurements the two systems no longer interact, no real change
can take place in the second system in consequence of anything that
may be done to the first system.''
\item {\it Reality}: ''If, without in any way disturbing a system, we can predict with certainty (i.e., with
probability equal to unity) the value of a physical quantity, then
there exists an element of physical reality corresponding to this
physical quantity.''
\item{\it Completeness}: ''Every element of the physical reality must have a counterpart in the [complete]
physical theory.''
\end{enumerate}
In contrast to the last three premises which, thought they are quite
plausible, are still indications of a certain philosophical
viewpoint, the first premise is a statement about a well established
property of a singlet state.

The EPR argument is as follows. Because of the perfect
anti-correlations (1.), we can predict with certainty the result of
measuring either $x$ component or $y$ component of spin of particle
2 by previously choosing to measure the same quantity of particle 1.
By locality (2.), the measurement of particle 1 cannot cause any
real change in particle 2. This implies that by the premise (3.),
both the $x$ {\it and} the $y$ components of spin of particle 2 are
elements of reality. This is also the case for particle 1 by a
parallel argument where particle 1 and 2 interchange their roles.
Yet, (according to Heisenberg's uncertainty principle) there is no
quantum state of a single spin in which both $x$ and $y$ spin
components have definite values. Therefore, by premise (4.) quantum
mechanics cannot be a complete theory.

In his answer~\cite{bohr1935}, published in the same year and under
the same title as of the EPR paper, Bohr criticized the EPR concept
of ''reality'' as assuming the systems having intrinsic properties
independently of whether they are observed or not and he argued for
''the necessity of a final renunciation of the classical ideal of
causality and a radical revision of our attitude towards the problem
of physical reality.'' Bohr pointed out that the wording of the
criterion of physical reality (3.) proposed by EPR contains an
ambiguity with respect to the expression "without in any way
disturbing the system". And, while, as Bohr wrote, there is ''no
question of mechanical disturbance of the system'', there is ''the
question of {\it an influence on the very conditions which define
the possible types of predictions regarding the future behavior of
the system.}'' Bohr thus pointed out that the results of quantum
measurements, in contrast to these of classical measurements, depend
on the complete experimental arrangement (context), which can even
be non-local as in the EPR case. Before any measurement is performed
only the correlations between the spin components of two particles,
but not spin components of individual particles are defined. The $x$
or $y$ component (but never both) of an individual particle becomes
defined only when the respective observable of the distant particle
is measured.

Perhaps the most clear way to see how strongly the philosophical
viewpoints of EPR and Bohr differ is in their visions of the future
development of quantum physics. While EPR wrote: ''We believe that
such [complete] a theory is possible'', Bohr's opinion is that (his)
complementarity ''provides room for new physical law, the coexistence
of which might at first sight appear irreconcilable with the basic
principles  of science.''

\section{Bell's theorem} \label{sec:1}

Bell's theorem can be thought of as a disproof of the validity of
EPR ideas. Elements of physical reality cannot be an internally
consistent notion. A broader interpretation of this result is that a
local and realistic description of nature, at the fundamental level,
is untenable. Further consequences are that there exist quantum
processes which cannot be medelled by any classical ones, not
necessarily physical processes, but also some classical computer
simulations with a communication constraint. This opened the
possibility of development of quantum communication.

We shall present now a derivation of Bell's inequalities. The stress
will be put on clarification of the underlying assumptions. These
will be presented in the most reduced form.

\subsection{Thought experiment}

At two measuring stations $A$ and $B$, which are far away from each
other, two characters Alice and Bob observe simultaneous flashy
appearances of numbers $+1$ or $-1$  at the displays of their local
devices (or the monitoring computers). The flashes appear in perfect
coincidence (with respect to a certain reference frame). In the
middle between the stations is something that they call ``source''.
When it is absent, or switched off, the numbers $\pm1$'s do not
appear at the displays. The activated source always causes two
flashes, one at $A$, one at $B$. They appear slightly after a
relativistic retardation time with respect to the activation of the
source, never before. Thus there is enough ``evidence'' for Alice
and Bob that the source causes the flashes. The devices at the
stations  have a knob which can be put in two positions: $m=1$ or
$2$ at $A $ station, and $n=1$ or $2$ at $B$. Local procedures used
to generate random choices of local knob positions are equivalent to
\emph{independent, fair coin tosses}. Thus, each of the four
possible values of the pair $n,m$  are equally likely, i.e. the
probability $P(n,m)=P(n)P(m)=\frac{1}{4}$. The ``tosses'', and knob
settings, are made at random times, and often enough, so that the
information on these is never available at the source during its
activation periods (the tosses and settings cannot have a causal
influence on the workings of source).  The local measurement data
(setting, result, moment of measurement) are stored and very many
runs of the experiment are performed.

\begin{figure}[t]
\sidecaption[t]
\includegraphics[scale=.6]{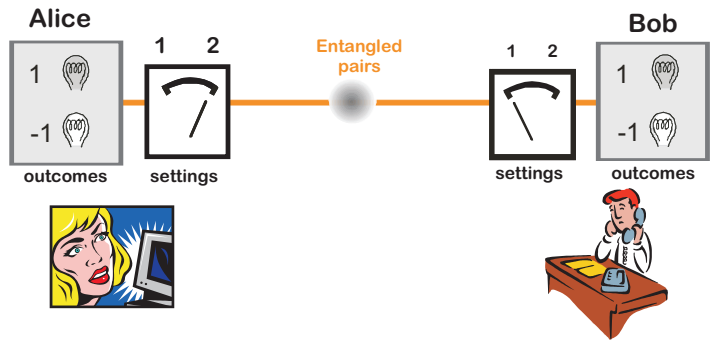}
\caption{Test of Bell's inequalities. Alice and Bob are two
separated parties who share entangled particles. Each of them is
free to choose two measurement settings 1 and 2 and they observe
flashes in their detection station which indicate one of the two
possible measurement outcomes +1 or -1.}
\label{fig:2}       
\end{figure}

\subsubsection{Assumptions leading to Bell's inequalities}

A concise {\em local realistic} description of such an experiment
would use the following assumptions \cite{GILL}:
\begin{enumerate}
\item
We assume {\em realism}, which is any logically self-consistent
model that allows one to use \emph{eight} variables in the
theoretical description of the experiment: $A_{m,n}$, $B_{n,m}$,
where $n,m=1,2$.  The variable $A_{m,n}$ gives the value, ${\pm 1}$,
which could be obtained at station $A$, if the knob settings, at $A$
and $B$, were at positions $n,m$, respectively. Similarly, $B_{n,m}$
plays the same role for station $B$, under the same settings. This
is equivalent to the assumption that a joint (non-negative, properly
normalized) probability distribution of these variables,
$P(A_{1,1},A_{1,2},A_{2,1},A_{2,2}; B_{1,1}, B_{1,2},B_{2,1}
B_{2,2})$, is always allowed to exist.\footnote{Note, that no hidden
variables appear, beyond these eight. However, given a (possibly
stochastic) hidden variables theory, one will be able to define our
eight variables as (possibly random) functions of the variables in
that theory.}
\item
The assumption of {\em locality} does not allow influences to go
outside the light cone.
\item Alice and Bob are free to choose their settings ``at the whim''. This the
\emph{freedom, or ``free will''}, often only a tacit
assumption~\cite{bellfree}. A less provocative version of this
assumption: {\em There exists stochastic processes which could be
used to choose the values of the local settings of the devices which
are independent of the workings of the source, that is they neither
influence it or are influenced by it.} By the previous assumptions
the events of activation of the source and of the choice and fixing
of the local settings must be space-like separated.
\end{enumerate}

Note that when setting labels $m$, $n$ are sent to the
measurement devices, they will likely cause some  unintended
disturbance: by these assumptions {\em any disturbance at A, as far as it influences the
outcome at A, is not related to the coin toss nor to the potential
outcomes at B, and vice versa}.

Note  further, that $A_{n,m}$ and $B_{n,m}$ are not necessarily
actual properties of the systems. The only thing that is assumed it
that there is a theoretical description which allows one to use
these all {\em eight} values.

\subsubsection{First consequences}
Let us write down the immediate consequences of these assumptions:

\begin{itemize}
\item By
{\em locality}: for all $n,m$:
\begin{equation}\label{e:locality}
A_{m,n}=A_{m},\quad B_{n,m}=B_n
\end{equation}
That is, the outcome which would appear at A does not depend on
which setting might be chosen at B, and vice versa. {\em Thus
$P(A_{1,1}, \ldots, B_{2,2})$ can be reduced to
$P(A_{1},A_{2},B_{1},B_{2})$.}
\item By {\em freedom}
\begin{equation}\label{e:freedom}
(n,m)~~\mbox{is statistically independent
of}~~(A_{1},A_{2},B_{1},B_{2}).\qquad
\end{equation}
\end{itemize}
Thus, the {\em overall} probability distributions for potential
settings and potential outcomes satisfy
\begin{equation}
P(n,m, A_{1},A_{2},B_{1},B_{2})=P(n,m)p(A_{1},A_{2},B_{1},B_{2})
\label{IDEPENDENCE}
\end{equation}
The choice of settings in the two randomizes,  $A$ and $B$, is
causally separated from the local realistic mechanism, which
produces the potential outcomes.

\subsubsection{Lemma: Bell's inequality}
The probabilities, $\Pr$, of the four logical propositions,
$A_n=B_m$, satisfy
\begin{equation}\label{e:Edelta}
\mathrm \Pr\{A_1=B_2\}
-\Pr\{A_1=B_1\}-\Pr\{A_2=B_1\}-\Pr\{A_2=B_2\}~\le~0.
\end{equation}
Proof: only four, or
two, or none of the propositions, in the left hand side of the
inequality can be true, thus (\ref{e:Edelta}). QED.

Now, if the observation settings are totally random (dictated by
``coin tosses''), $P(n,m)=\frac{1}{4}$. Then, according to all our
assumptions  \BEQ P(A_n=B_m| n,m)=
P(n,m)\Pr\{A_n=B_n\}=\frac{1}{4}\Pr\{A_n=B_m\}.\EEQ Therefore, we
have a  Bell inequality: under the {\em conjunction} of the
assumptions for the {\it experimentally accessible} probabilities
one has
\begin{equation}\label{e:bellinequ}
P(A_1=B_2\mid 1,2)-P(A_1=B_1\mid 1,1) -P(A_2=B_1\mid
2,1)-P(A_2=B_2\mid 2,2)~\le~0.
\end{equation}
This is the well-known Clauser-Horne-Shimony-Holt (CHSH)
inequality~\cite{chsh}.

\subsection{The Bell theorem} Quantum mechanics predicts for some
experiments satisfying all the features of the thought experiment
the left hand side of inequality (\ref{e:bellinequ}) to be as high
as $\sqrt 2 -1,$ which is larger that the local realistic bound 0.
{\em Hence, one has Bell's theorem~\cite{bell1964}: if quantum
mechanics holds, local realism, defined by the full set of the above
assumptions, is untenable.} But, how does nature behave -- according
to local realism or quantum mechanics? It seems that we are
approaching the moment, in which one could have as perfect as
possible laboratory realization of the thought experiment (locality
loophole was closed in~\cite{WEIHS,aspect}, detection loophole
in~\cite{rowe} and in recent experiment measurement settings were
space-like separated from the photon pair emission~\cite{scheidl}).
Hence local realistic approach to description of physical phenomena
is close to be shown untenable too.

\subsubsection{The assumptions as a communication complexity problem}

Assume that we heave two programmers $P_k$, where $k=1,2$, each
possessing an enormously powerful computer. They share certain joint
classical information strings of arbitrary lengths and/or some
computer programs. All these will be collectively denoted as
$\lambda$.  But, once they both posses $\lambda$, no communication
whatsoever between them is allowed. After this initial stage, each
one of them gets from a Referee a one bit random number $x_k \in
\{0,1\}$, known only to him/her ($P_1$ knows only $x_1$, $P_2$ knows
only $x_2$). The {\em individual} task of each of them is to
produce, via whatever computational  program, a one bit number
$I_k(x_k, \lambda)$, and communicate only this one bit to a Referee,
who just compares the received bits. There is no restriction on the
form and complication of the {\em possibly stochastic}
functions $I_k$, or any actions taken to define the values, but any
communication between the partners is absolutely not allowed. The
{\em joint} task of the partners is to devise a computer code which
under the constraints listed above, and without any cheating, allows
to have after very many repetitions of the procedures (each starting
with establishing a new shared $\lambda$) the following functional
dependence of the probability that their bits sent back to the
Referee are equal:
\begin{equation}
P\{I_1(x_1)=I_2(x_2)\}=\frac{1}{2}+
\frac{1}{2}\cos\big[-\pi/4+(\pi/2)(x_1+x_2)\big].
\label{QUANTUM-PROB}
\end{equation}
This is a variant of communication complexity problems. The current
task is absolutely impossible to achieve with the classical means at
their disposal, and without communication. Simply because whatever
is the protocol
\begin{equation}\label{e:bellinequ1}
\Pr\{I_1(1)=I_2(1)\}-\Pr\{I_1(0) =I_2(0)\}
-\Pr\{I_1(1)=I_2(0)\}-\Pr\{I_1(0)=I_2(1)\} ~\le~0.
\end{equation}
whereas, the  value of this expression in quantum strategy $P_Q$ can
be as high as $\sqrt{2}-1$. If the programmers use entanglement as
resource and receive their respective qubits from an entangled pair
(e.g. singlet) during the communication stages (when $\lambda$ is
established), one can obtain on average $P_Q$. Instead of computing,
the partners make a local measurement on their qubits. They measure
Pauli observables $\vec{n}\cdot\vec{\sigma}$, where $||\vec{n}||=1$.
Since the probability for them to get identical results, $r_1,r_2$,
for observation directions $\vec{n}_1,\vec{n}_2$ is
\begin{equation}
P_Q\{r_1=r_2|\vec{n}_1,\vec{n}_2\}=\frac{1}{2}-
\frac{1}{2}\vec{n}_1\cdot \vec{n}_2,
\end{equation}
for suitably chosen $\vec{n}_1(x_1),\vec{n}_2(x_2)$ they get values
of $P_Q$ equal to those in (\ref{QUANTUM-PROB}). The messages sent
back to the Referee encode the local results of measurements of
$\vec{n_1}\cdot\vec{\sigma} \otimes \vec{n_2}\cdot\vec{\sigma}$, and
the local measurement directions are suitably chosen as functions of
$x_1$ and $x_2$. We will come back to the relation between Bell's
inequalities and quantum communication complexity problems in more
details in Sec.~\ref{secqcc}.

\subsubsection{Philosophy or physics? Which assumptions?}

The assumptions behind Bell inequalities are often criticized as
being ``philosophical''. If one reminds oneself on Mach's influence
on Einstein, philosophical discussions related to physics may be
very fruitful.

For those who are, however, still skeptical one can argue as
follows. The whole (relativistic) classical theory of physics is
realistic (and local). Thus we have an important exemplary
realization of the postulates of local realism. Philosophical
propositions could be defined as those which {\em are not}
observationally or experimentally falsifiable at the given moment of
the development of human knowledge, or in pure mathematical theory
are not logically derivable. Therefore, the {\em conjunction } of
all assumptions of Bell inequalities is not a philosophical
statement, as it is {\em testable} both experimentally and logically
(within, known at the moment, mathematical formulation of
fundamental laws of physics). Thus, Bell's theorem  removed the
question of possibility of local realistic description from the
realm of philosophy. Now this is just a question of a good
experiment.

The other criticism is formulated in the following way. Bell
inequalities can be derived using a single assumption of existence
of joint probability distribution for the observables involved in
them, or that the probability calculus of the experimental
propositions involved in the inequalities is of Kolmogorovian
nature, and nothing more. But if we want to apply these assumptions
to the thought experiment we stumble on the following question: {\em
does the joint probability take into account full experimental
context or not?}. The experimental context is in our case (at least)
the full state of the settings $(m,n)$. Thus if we use the same
notation as above for the realistic values, this time applied to the
possible results of measurements of observables, initially we can
assume existence of only $p(A_{1,1},A_{1,2},A_{2,1},A_{2,2};
B_{1,1}, B_{1,2},B_{2,1}, B_{2,2})$. Note that such a probability
could be e.g. factorizable into $\prod_{n,m}P(A_{n,m},B_{n,m})$.
That is one could in such a case have different probability
distributions pertaining to different experimental contexts (which
can even be defined through the choice of measurement settings in
space-like separated laboratories!)

Let us discuss this from the quantum mechanical point of view, only
because such considerations have a nice formal description within
this theory, familiar to all physicists. Two observables, say
$\hat{A}_1\otimes\hat{B}_1$ and $\hat{A}_2\otimes\hat{B}_2$, as well
as other possible pairs are functions of two different {\em maximal}
observables for the whole system (which are non-degenerate by
definition). If one denotes such a maximal observable linked with
$\hat{A}_m\otimes\hat{B}_n$ by $\hat{M}_{m,n}$ and its eigenvalues
by $M_{m,n}$ the existence of the aforementioned joint probability
is equivalent to the existence of a
$p(M_{1,1},M_{1,2},M_{2,1},M_{2,2})$ in form of a proper probability
distribution. Only if one assumes additionally context independence,
this can be reduced to the question of existence of (non-negative)
probabilities $P(A_{1},A_{2},B_{1},B_{2})$, where $A_{m}$ and
$B_{n}$ are eigenvalues of $\hat{A}_m\otimes\mathbf{1}$ and
$\mathbf{1}\otimes\hat{B}_n$, where it turn $\mathbf{1}$ is the unit
operator for the given subsystem. While context independence is
physically doubtful, when the measurements are not spatially
separated, and thus one can have mutual causal dependence, it is
well justified for spatially separated measurements. I.e., {\em
locality} enters our reasoning, whether we like it or not. Of course
one cannot derive any Bell inequality of the usual type if the
random choice of settings is not independent of the distribution of
$A_{1},A_{2},B_{1},B_{2}$, that is without (\ref{IDEPENDENCE}).

There is yet another challenge to the set of assumptions presented
above. It is often claimed, that realism can be derived, once one
considers the fact that maximally  entangled quantum systems reveal
perfect correlations, and one additionally assumes locality.
Therefore it would seem that the only basic assumption behind Bell
inequalities is locality, with the other auxiliary ones of freedom.
Such a claim is based on the ideas of EPR, who conjectured that one
can introduce ``elements of reality'' of a remote system, provided
this system is perfectly correlated with another system. To show the
fallacy of such a hope, let us now discuss three particle
correlations, in the case of which consideration of just few
``elements of reality'' reveals that they are a logically
inconsistent notion. Therefore, they cannot be a starting point for
deriving a self-consistent realistic theory. The three particle
reasoning is used here because of its beauty and simplicity, not
because one cannot reach a similar conclusion for two particle
correlations.

\subsection{Bell's theorem without inequalities: three entangled particles or more}

As the simplest example, take a
Greenberger-Horne-Zeilinger~\cite{ghz} (GHZ) state of $N=3$
particles (fig.2):
\begin{equation}
|\mbox{GHZ}\rangle = \frac{1}{\sqrt{2}}\left(|a\rangle |b\rangle
|c\rangle +|a'\rangle |b'\rangle |c'\rangle\right) \label{1}
\end{equation}
where $\langle x|x'\rangle=0$ ($x=a,b,c$, and kets denoted by one
letter pertain to one of the particles). The observers, Alice, Bob
and Cecil  measure the observables: $\hat{A}(\phi_A)$,
$\hat{B}(\phi_B)$, $\hat{C}(\phi_C)$, defined by
\begin{equation}
\hat{X}(\phi_X)
 = |+,\phi_X\rangle  \langle+,\phi_X |
- |-,\phi_X\rangle  \langle-,\phi_X | \label{2}
\end{equation}
where
\begin{equation}
 |\pm,\phi_X\rangle
= \frac{1}{\sqrt{2}}\left(\pm i|x'\rangle + \exp{(i\phi_X
)}|x\rangle\right) . \label{3}
\end{equation}
and $\hat{X}=\hat{A},\hat{B},\hat{C}$. The quantum prediction for
the expectation value of the product of the three local observables
is given by
\begin{equation}
E(\phi_A ,\phi_B ,\phi_c) = \langle
\mbox{GHZ}|\hat{A}(\phi_A)\hat{B}(\phi_B)\hat{C}(\phi_C)
|\mbox{GHZ}\rangle = \sin(\phi_A +\phi_B+\phi_c).
\end{equation}
Therefore, if $\phi_A +\phi_B+\phi_c=\pi/2+k\pi$, quantum mechanics
predicts perfect correlations. For example, for $\phi_A=\pi/2$,
$\phi_B=0$ and $\phi_c=0$, whatever may be the results of local
measurements of the observables, for say the particles belonging to
the $i$-th triple represented by the quantum state
$|\mbox{GHZ}\rangle$, their product must be unity. In a local
realistic theory one would have
\begin{equation}
A^i(\pi/2) B^i(0) C^i(0) = 1, \label{4}
\end{equation}
where $X^i(\phi)$, $X=A,B$ or $C$ is the local realistic value of a
local measurement of the observable $\hat{X}(\phi)$ that {\it would
have been} obtained for the $i$-th particle triple if the setting of
the measuring device  is  $\phi$. By locality $X^i(\phi)$ depends
solely on the local parameter. The eq. (\ref{4}) indicates that we
can predict with certainty the result of measuring the observable
pertaining to one of the particles (say $c$) by choosing to measure
suitable observables for the other two. Hence the value $X^i(\phi)$
are EPR elements of reality.

\begin{figure}[t]
\sidecaption[t]
\includegraphics[scale=.47]{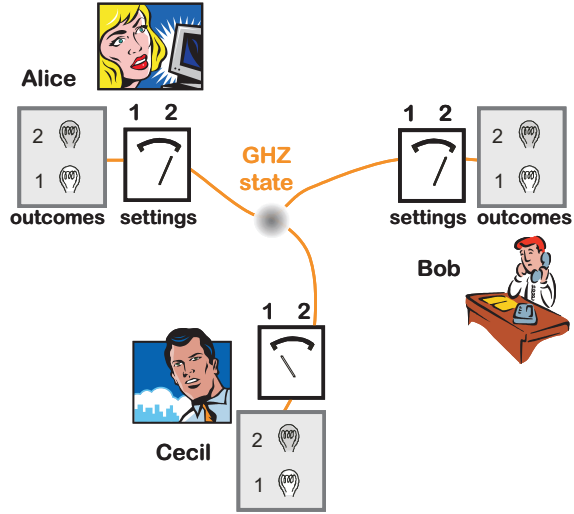}
\caption{Test of the GHZ theorem. Alice, Bob and Cecil are three
separated parties who share three entangled particles in the GHZ
state. Each of them are free to choose between two measurement
settings 1 and 2 and they observe flashes in their detection station
which indicate one of the two possible measurement outcomes +1 or
-1.}
\label{fig:2}       
\end{figure}

However, if the local apparatus settings are different one {\it
would have had}, e.g.
\begin{eqnarray}
&A^i(0) B^i(0) C^i(\pi/2) = 1,&\\ \label{5}
 &A^i(0) B^i(\pi/2)
C^i(0) = 1,&\\ \label{6}
 &A^i(\pi/2) B^i(\pi/2) C^i(\pi/2) = -1.&
\label{7}
\end{eqnarray}
Yet, the four statements (\ref{4}-\ref{6}) are inconsistent within
local realism. Since $X^i(\phi)=\pm1$, if one multiples side by side
the eqs. (\ref{4}-\ref{6}), the result is
\begin{equation}
1= -1.
\label{8}
\end{equation}
This shows that the mere concept of existence of "elements of
physical reality" as introduced by EPR is in a contradiction with
quantum mechanical predictions. We have a ``Bell's theorem without
inequalities''~\cite{ghz}.

Some people still claim that EPR correlations together with the
assumption of locality allow one to derive realism. The above
example clearly shows that such a realism would allow one to infer
that $1=-1$.

\subsection{Implications of Bell's theorem}

Violations of Bell's inequalities
imply that the underlying {\em conjunction of assumptions of
realism, locality and ``free will''} is not valid, and {\em nothing
more}.

It is often said that the violations indicate ``(quantum)
non-locality''. However if one  wants {\em non-locality} to be {\em
the} implication, one has to assume ``free will'' and realism. But
this is only at this moment a philosophical choice (it seems that
there is no way to falsify it). {\em It is not a necessary condition
for violations of Bell's inequalities.}

The theorem of Bell shows that even a local inherently probabilistic
hidden-variable theory cannot agree with all predictions of quantum
theory (we base our considerations on $p(A_1,A_2,B_1,B_2)$ without
assuming its actual structure, or whether the distribution for a
single run is essentially deterministic, all we require is a joint
``co-existence'' of the variables $A_1,...,B_2$ in a {\em
theoretical} description). Therefore the above statements cover
theories that treat probabilities as irreducible, and for which one
can define $p(A_1,A_2,B_1,B_2)$. Such theories contradict quantum
predictions. This, for some authors indicates that nature is
non-local. While the mere existence of Bohm's model~\cite{bohm}
demonstrates that non-local hidden-variables are a logically valid
option, we now know that there are plausible models, such as
Leggett's crypto-nonlocal hidden-variable model~\cite{leggett}, that
are in disagreement with both quantum predictions and
experiment~\cite{groablacher}. But, perhaps more importantly, if one
is ready to consider inherently probabilistic theories, then there
is no immediate reason to require the existence of (non-negative and
normalized) probabilities $p(A_{1,1}...,B_{2,2})$. Violation of this
condition on realism, together with locality, which allows one to
reduce the distribution to $p(A_1,..., B_2)$, is not in a {\em
direct} conflict with the theory of relativity, as it does not
necessarily imply the possibility of signalling superluminally. To
the contrary, quantum correlations cannot be used for direct
communication between Alice to Bob, but still violate Bell's
inequalities. It is therefore legitimate to consider quantum theory
as a probability theory subject to, or even derivable from more
general principles, such as non-signaling
condition~\cite{popescurohrlich,barrett} or information theoretical
principles~\cite{weizsaecker,zeilinger}.

Note that complementarity, inherent in quantum
formalism\footnote{Which can be mathematically expressed as
non-existence of joint probabilities for non-commuting, i.e.
non-commeasurable, observables.}, completely contradicts the form of
realism defined above. So why quantum-non-locality?

To put it short, Bell's theorem does not imply {\em any} property of quantum mechanics.
It just tells what it is not.

\section{All Bell's inequalities for two possible settings on each
side}

We shall now present a general method of deriving {\em all} standard
Bell inequalities (that is Bell's inequalities involving two-outcome
measurements and with two settings per observer). Although these
will not be spelled out explicitly, all the assumptions discussed
above are behind the algebraic manipulations leading to the
inequalities. We present in detail a derivation for two-observer
problem, because the generalization to more observers is,
surprisingly, obvious.

Consider pairs of particles (say, photons) simultaneously  emitted
in well defined opposite directions. After some time the photons
arrive  at two  very distant measuring devices A and B operated by
Alice and Bob. Alice, chooses to measure  either observable
$\hat{A}_1$ or $\hat{A}_2$, and Bob either $\hat{B}_1$ or
$\hat{B}_2$. The hypothetical results that they may get for the
$j$-th pair of photons are $A_1^j$ and  $A_2^j$, for Alice's two
possible choices, and $B_1^j$ and $B_1^j$, for Bob's. The numerical
values of these results ($+1$ or $-1$) are defined by the two
eigenvalues of the observables.

Since, always either $|B_1^j -B_2^j|=2$ and $|B_1^j +B_2^j|=0$, or
$|B_1^j -B_2^j|=0$ and $|B_1^j +B_2^j|=2$, with a similar property
of Alice's hypothetical results the following relation holds
\begin{eqnarray}
|A_1^j \pm A_2^j| \cdot |B_1^j \pm B_2^j|=0\label{relation}
\end{eqnarray}
for all possible sign choices within (\ref{relation}) except one,
for which one has $4$. Therefore
\begin{eqnarray}
\sum_{k,l=0}^{1}|(A_1^j +(-1)^k A_2^j)(B_1^j+(-1)^lB_2^j)|=4,
\label{BELLEQ}
\end{eqnarray}
or equivalently one has the set of identities
\begin{eqnarray}
\sum_{s_1,s_2=-1}^{1} S(s_1,s_2)[(A_1^j +s_1
A_2^j)(B_1^j+s_2B_2^j)]=\pm4, \label{BELLEQ}
\end{eqnarray}
with any $S(s_1,s_2)=\pm1$. There are $2^{2^2}=16$ such $S$
functions.

Imagine now that  $N$ pairs of photons  are emitted, pair by pair
($N$ is sufficiently large, such that $\sqrt{1/N}\ll 1$). The
average value of the products of the local values is given by
\begin{equation}
E(A_n,B_m)=\frac{1}{N}\sum_{j=1}^{N}A_n^jB_m^j,
\end{equation}
where $n,m=1,2$.

Therefore after averaging,  the following  single Bell-type
inequality emerges:
\begin{equation}
\sum_{k,l=0}^{1} |E(A_1,B_1)+(-1)^lE(A_1,B_2) +(-1)^k E(A_2,B_1)
+(-1)^{k+l}E(A_2,B_2)|\leq4,\label{BELLINEQ}
\end{equation}
or equivalently a series of inequalities:
\begin{equation}
\sum_{s_1,s_2=-1}^{1}
S(s_1,s_2)[E(A_1,B_1)+s_2E(A_1,B_2)+s_1E(A_2,B_1)
+s_1s_2E(A_2,B_2)]\leq4. \label{BELLINEQ-2}
\end{equation}
As the choice of measurement settings is assumed to be statistically
independent of the working of the source, i.e of the distribution of
$A_1$'s, $A_2$'s, $B_1$'s and $B_2$'s, the averages $E(A_n,B_m)$
cannot differ much, for high $N$, from the {\it actually observed}
ones in the subsets of runs for which the given pair of settings was
selected.

\subsection{Completeness of the inequalities}

The inequalities form a complete set. That is, they
define the faces of the convex polytope formed out of all possible local realistic models for the given set of measurements.
Whenever local realistic model exists inequality~(\ref{BELLINEQ}) is
satisfied by its predictions. To prove the sufficiency of
condition~(\ref{BELLINEQ}) we construct a local realistic model for
any correlation functions which satisfy it, i.e. we are interested
in the local realistic models for $E_{k_1k_2}^{LR}$ such that they
fully agree with the measured correlations $E(k_1,k_2)$ for all
possible observables $k_1,k_2=1,2$.

One can introduce $\hat E$ which is a ``tensor" or matrix built out
of $E_{ij}$, with $i,j=1,2$. If all its components can be derived
from local realism, one must have
\begin{equation}
\hat E_{LR} = \sum_{\vec A,\vec B =-1}^{1} P(\vec A,\vec B) \vec A
\otimes \vec B,
\end{equation}
with $\vec A = (A_1(\vec{n}_{1}), s_1 A_1(\vec{n}_{2}))$, $\vec B =
(A_2(\vec{n}_{1}), s_2 A_2(\vec{n}_{2}))$, where $s_1,s_2\in
\{-1,1\}$ and nonnegative normalized probabilities $P(\vec A,\vec
B)$.

Let us ascribe for fixed $s_1,s_2$, a hidden probability that
$A_j(\vec{n}_{1}) = s_j A_j(\vec{n}_{2})$ (with $j=1,2$) in the form
familiar from Eq.~(\ref{BELLINEQ}):
\begin{equation}
P(s_1,s_2)=\frac{1}{4} |\sum_{k_2,k_2=1}^{2} s_1^{k_1-1} s_2^{k_2-1}
E(k_1,k_2)|. \label{PROB}
\end{equation}
Obviously these probabilities are positive. However they sum up to
identity only if inequality~(\ref{BELLINEQ}) is saturated, otherwise
there is a ``probability deficit'', $\Delta P$. This deficit can be
compensated without affecting correlation functions.

First we construct the following structure, which is indeed the
local realistic model of the set of correlation functions if the
inequality is saturated:
\begin{equation}
\sum_{s_1,s_2=-1}^{1} \Sigma(s_1,s_2) P(s_1,s_2) (1,s_1) \otimes
(1,s_2), \label{MODEL}
\end{equation}
where $\Sigma(s_1,s_2)$ is the sign of the expression within the
modulus in Eq. (\ref{PROB}).

Now if $\Delta P > 0$, we add a ``tail'' to this expression given
by:
\begin{equation}
\frac{\Delta P}{16} \sum_{A_1=-1}^{1} \sum_{A_2=-1}^{1}
\sum_{B_1=-1}^{1} \sum_{B_2=-1}^{1} (A_1,A_2) \otimes (B_1,B_2).
\end{equation}
This ``tail'' does not contribute to the values of the correlation
functions, because it represents the fully random noise. The sum of
(\ref{MODEL}) is a valid local realistic model for $\hat E =
(E(1,1),E(1,2),E(2,1),E(2,2))$. The sole role of the ``tail'' is to
make all hidden probabilities to add up to $1$.

%

To give the reader some intuitive grounds for the actual form of,
and the completeness of the derived inequalities, we shall now give
some remarks. The gist is that the consecutive terms in the
inequalities are just expansion coefficients of the tensor $\hat{E}$
in terms of a complete orthogonal sequence of basis tensors. Thus
the expansion coefficients represent the tensors in a one-to-one
way.

In the four dimensional real space where both ${\hat E}_{LR}$ and
$\hat E$ are defined one can find an orthonormal basis set $\hat
S_{s_1s_2} = \frac{1}{2}(1,s_1)\otimes(1,s_2)$. Within these
definitions the hidden probabilities acquire a simple form:
\begin{equation}
P(s_1,s_2) = \frac{1}{2} |\hat S_{s_1s_2} \cdot \hat E|,
\end{equation}
where the dot denotes the scalar product in $\mathbf{R}^4$. Now the
local realistic correlations, $\hat E_{LR}$, can be expressed as:
\begin{equation}
\hat E_{LR} = \sum_{s_1,s_2=-1}^{1} |\hat S_{s_1s_2} \cdot \hat E|
\Sigma(s_1,s_2) \hat S_{s_1s_2}.
\end{equation}
The modulus of any number $|x|$ can be split into $|x| = x \textrm{
sign}(x)$, and we can always demand the product $A_1(\vec{n}_{1})
A_2(\vec{n}_{1})$ to have the same sign as the expression inside the
modulus. Thus we have:
\begin{equation}
\hat E = \sum_{s_1,s_2=-1}^{1} (\hat S_{s_1s_2} \cdot {\hat E}) \hat
S_{s_1s_2}.
\end{equation}
The expression in the bracket is the coefficient of tensor $\hat E$
in the basis $\hat S_{s_1s_2}$. These coefficients are then summed
over the same basis vectors, therefore the last equality appears.

\subsection{Two-qubit states that violate the inequalities}

A general two qubit state can be put in the following concise form
\begin{equation}
 \hat{\rho}=\frac{1}{4} \sum_{\mu,\nu=0}^{3} T_{\mu\nu}
(\hat{\sigma}^1_{\mu}\otimes \hat{\sigma}^2_{\nu}).
\end{equation}
The two qubit correlation function for measurements of spin 1 along
direction $\vec{n}(1)$ and of spin 2 along $\vec{n}(2)$ is given by
\begin{equation}
E_{QM}(\vec{n}(1),\vec{n}(2))=\mbox{Tr}\left[
 \hat{\rho}\left(\vec{n}(1)\cdot{\vec{\hat{\sigma}}}^1\otimes
\vec{n}(2)\cdot {\vec{\hat{\sigma}}}^2 \right)\right], \label{23D}
\end{equation}
and it reads
\begin{equation}
E_{QM}(\vec{n}(1),\vec{n}(2))=
 \sum_{i,j=1}^{3} T_{ij}{n}(1)_i{n}(2)_j.
\label{24}
\end{equation}
Two particle correlations are fully defined once one knows the
components of $ T_{ij}$, $i,j=1,2,3$, of the tensor ${\bf \hat{T}}$.
Equation (\ref{24}) can be put into a more convenient form:
\begin{equation}
E_{QM}(\vec{n}(1),\vec{n}(2))= {\bf \hat{T}} \bullet
\vec{n}(1)\otimes\vec{n}(2), \label{25}
\end{equation}
where ''$\bullet$'' is the scalar product in the space of tensors,
which in turn is isomorphic with $\mathbf{R}^3\otimes \mathbf{R}^3$.

Quantum correlation $E_{QM}(\vec{n}(1),\vec{n}(2))$ can be described
by a local realistic model if, and only if, for {\it any} choice of
the settings $\vec{n}(1)^{k_1}$ and $\vec{n}(2)^{k_2}$, where
$k_1,k_2=1,2$, one has
\begin{equation}
\frac{1}{4} \sum_{k,l=1}^{2} \left| {\bf \hat{T}} \bullet
[\vec{n}(1)^{1}+(-1)^k\vec{n}(1)^{2}]\otimes[\vec{n}(2)^{1}
+(-1)^l\vec{n}(2)^{2})] \right| \leq 1. \label{27}
\end{equation}
Since there always exist two mutually orthogonal unit  vectors
$\vec{a}(x)^{1}$ and  $\vec{a}(x)^{2}$ such that
\begin{equation}
\vec{n}(x)^{1}+(-1)^k\vec{n}(x)^{2}= 2\alpha(x)_k\vec{a}(x)^{k}
\mbox{ with } k=1,2
\end{equation}
and with $\alpha(x)_1=\cos{\theta(x)}$,
$\alpha(x)_2=\sin{\theta(x)}$, one obtains
\begin{equation}
\sum_{k,l=1}^{2} \left| \alpha(1)_k\alpha(2)_l {\bf \hat{T}} \bullet
\vec{a}(1)^{k}\otimes\vec{a}(2)^{l} \right| \leq 1. \label{27A}
\end{equation}
Note that ${\bf \hat{T}}\bullet\vec{a}(1)^{k}\otimes\vec{a}(2)^{l}$
is a component of the tensor ${\bf \hat{T}}$ after a transformation
of the local coordinate systems of each of the particles into such
ones where the two first basis vectors are $\vec{a}(x)^{1}$ and
$\vec{a}(x)^{2}$. We shall denote such transformed components again
by $T_{kl}$.

The necessary and sufficient condition for a two-qubit correlation
to be described within a local realistic model is that in any plane
of observations for each particle (defined by the two observation
directions) one must have
\begin{equation}
\sum_{k,l=1}^{2} \left| \alpha(1)_k\alpha(2)_l T_{kl}\right| \leq 1.
\label{27B}
\end{equation}
for arbitrary $\alpha(1)_k$, $\alpha(2)_l$.

Using the Cauchy inequality one obtains
\begin{equation}
\sum_{k,l=1}^{2} \left| \alpha(1)_k\alpha(2)_lT_{kl}\right|\leq
\sqrt{ \sum_{k,l=1}^{2} T_{kl}^2}. \label{27C}
\end{equation}

Therefore, if
\begin{equation}
\sum_{k,l=1}^{2}T_{kl}^2 \leq 1 \label{28}
\end{equation}
for any set of local coordinate systems, the two particle
correlation functions of the form of (\ref{24}) can be understood
within the local realism (in a two settings per observer
experiment).

{\em This condition is both necessary and sufficient.}

\subsubsection{Sufficient condition for violation of the inequality}

The full set of inequalities is derivable from the
identity~(\ref{BELLEQ}) where we put non-factorable sign function
$S(s_1,s_2)=\frac{1}{2}(1+s_1)+(1-s_1)s_2$. In this case one obtains
the CHSH inequality in its standard form:
\begin{equation}
\left| \left\langle (A_1+ A_2)B_1 + (A_1 - A_2)B_2
\right\rangle_{avg} \right| \le 2, \label{full22}
\end{equation}
where $\langle ... \rangle_{arg}$ denotes average. All other
non-trivial inequalities are obtainable by all possible sign changes
$X_k \to -X_k$ (with $k=1,2$ and $X = A,B$). It is easy to see that
factorizable sign functions, such as e.g. $S(s_1,s_2)=s_1s_2$, lead
to trivial inequalities $|E(A_n, B_m)|\leq 1$. As noted above the
quantum correlation function  $E_Q(\vec a_k,\vec b_l)$ is given by
the scalar product of the correlation tensor ${\bf \hat T}$ with the
tensor product of the local measurement settings represented by unit
vectors $\vec a_k \otimes \vec b_l$, i.e. $E_Q(\vec a_k,\vec b_l) =
(\vec a_k \otimes \vec b_l) \cdot {\bf \hat T}$. Thus, the condition
for a quantum state endowed with the correlation tensor ${\bf \hat
T}$ to satisfy the inequality (\ref{full22}), is that for all
directions $ \vec a_1, \vec a_2, \vec b_1, \vec b_2$ one has
\begin{equation}
\Big| \Big[ (\frac{\vec a_1 + \vec a_2}{2}) \otimes \vec b_1 +
(\frac{\vec a_1 - \vec a_2}{2}) \otimes \vec b_2 \Big] \cdot {\bf
\hat T} \Big| \le 1, \label{QUANTUM_INEQ}
\end{equation}
where both sides of (\ref{full22}) were divided by $2$.

Next notice that $\vec A_{\pm} = \frac{1}{2}(\vec a_1 \pm \vec a_2)$
satisfy the following relations: $\vec A_+ \cdot \vec A_- = 0$ and
$||\vec A_+ ||^2 + ||\vec A_- ||^2 = 1$. Thus $\vec A_+ + \vec A_-$
is a unit vector, and $\vec A_\pm$ represent its decomposition into
two orthogonal vectors. If one introduces unit vectors $\vec a_\pm$
such that $\vec A_\pm = a_\pm \vec a_\pm$, one has $a_+^2 + a_-^2 =
1$. Thus one can put inequality (\ref{QUANTUM_INEQ}) into the
following form:
\begin{equation}
|{\bf \hat S} \cdot {\bf \hat T}| \le 1,
\end{equation}
where ${\bf \hat S} = a_+ \vec a_+ \otimes \vec b_1 + a_- \vec a_-
\otimes \vec b_2$. Note that since $\vec a_+ \cdot \vec a_- = 0$,
one has ${\bf \hat S} \cdot {\bf \hat S} = 1$, i.e. ${\bf \hat S}$
is a tensor of unit norm. Any tensor of unit norm, ${\bf \hat U}$,
has the following Schmidt decomposition ${\bf \hat U} = \lambda_1
\vec v_1 \otimes \vec w_1 + \lambda_2 \vec v_2 \otimes \vec w_2$,
where $\vec v_i \cdot \vec v_j = \delta_{ij}, \vec w_i \cdot \vec
w_j = \delta_{ij}$ and $\lambda_1^2 + \lambda_2^2 = 1$. The
(complete) freedom of  the choice of the measurement directions
$\vec b_1$ and $\vec b_2$, allow one by choosing $\vec b_2$
orthogonal to $\vec b_1$ to put ${\bf \hat S}$ in the form
isomorphic with ${\bf \hat U}$, and the freedom of choice of $\vec
a_1$ and $\vec a_2$ allows $\vec A_+$ and $\vec A_-$ to be arbitrary
orthogonal unit vectors, and $\vec a_+$ and $\vec a_-$ to be also
arbitrary. Thus ${\bf \hat S}$ can be equal to any unit tensor. To
get the maximum of the left hand side of (\ref{QUANTUM_INEQ}), we
Schmidt decompose the correlation tensor, and take two terms of the
decomposition which have the largest coefficients. In this way we
get a tensor ${\rm \hat{T}'}$, of Schmidt rank two. We put ${\rm\hat
S} = \frac{1}{||{\rm \hat T'}||} {\rm\hat T'}$, and the maximum is
$||{\rm \hat T'}|| = \sqrt{{\rm \hat T'} \cdot {\rm \hat T'}}$.
Thus, in other words,
\begin{equation}
\max\big[\sum_{k,l=1}^{2} T_{kl}^2 \big] \le 1 \label{t2-22}
\end{equation}
is the necessary and sufficient condition for the
inequality~(\ref{BELLINEQ}) to hold, provided the maximization is
taken over all local coordinate systems of two observers. {The
condition is equivalent to the necessary and sufficient condition of
Horodeccy Family ~\cite{horodecki} for violation of the CHSH
inequality.}

\subsection{Bell's inequalities for N particles}

Let us consider a Bell inequality test with $N$ observers. Each of
them chooses between two possible observables, determined by local
parameters  $\vec{n}_1(j)$ and $\vec{n}_2(j)$, where $j=1,...,N$.
Local realism implies existence of two numbers $A^j_1$ and $A^j_2$,
each taking values +1 or -1, which describe the predetermined result
of a measurement by the $j$-th observer for the two observables. The
following algebraic identity holds:
\begin{equation}
\sum_{s_1,...,s_N =-1}^{1} S(s_1,...,s_N) \prod_{j=1}^N [ A^j_1 +
s_j A^j_2]=\pm 2^N, \label{INEQ2}
\end{equation}
where $S(s_1,...,s_N)$ is an arbitrary "sign" function, i.e.
$S(s_1,...,s_N)=\pm 1$. It is a straightforward generalization of
the one for two observers as given in~(\ref{BELLINEQ-2}).
 The correlation function is the average over many runs of the experiment $
E_{k_1,...,k_N}=\langle  \prod_{j=1}^N A^j_{k_j} \rangle_{avg}$ with
$k_1,...k_N\in \{1,2\}$. After averaging (\ref{INEQ2}) over the
ensemble of the runs one obtains the Bell inequalities
\footnote{This set of inequalities is a sufficient and necessary
condition for the correlation functions entering them to have a
local realistic model. Compare it to the two particle case.}
\begin{equation}
|\sum_{s_1,...,s_N=-1}^{1} S(s_1,...,s_N) \sum_{k_1,...,k_N = 1}^{2}
s^{k_1-1}_1... s^{k_N-1}_N E_{k_1,...,k_N}| \leq 2^N.
\end{equation}
Since there are $2^{2^N}$ different functions $S$, the above
inequality represents a set of $2^{2^N}$ Bell inequalities.

All these boil down to just one inequality (!):
\begin{equation}
\sum_{s_1,...,s_N = -1}^{1} |\sum_{k_1,...,k_N= 1}^{2} s^{k_1-1}_1
... s^{k_N-1}_N E_{k_1,...,k_N}| \leq 2^N,
\end{equation}
The proof of this fact is trivial exercise with the use of the
property that either $|X|=1$ or $|X|=-1$, where $X$ is a real
number. This inequality was derived independently in Refs \cite{WZ}
and ~\cite{wernerwolf}. The presented derivation follows mainly
Ref.~\cite{zukowskibrukner}.

\subsection{N-qubit correlations}

A general N-qubit state can be put in the form
\begin{equation}
\hat{\rho}=\frac{1}{2^N} \sum_{\mu_1,\cdots, \mu_N=0}^{3}
T_{\mu_1\cdots\mu_N} (\otimes_{k=1}^N \hat{\sigma}^k_{\mu_K}).
\end{equation}
Thus, the $N$ qubit correlation function has the following structure
\begin{equation}
E_{QM}(\vec{n}(1),\vec{n}(2),...,\vec{n}(N))= {\bf \hat{T}} \bullet
\vec{n}(1)\otimes\vec{n}(2)...\otimes\vec{n}(N), \label{25N}
\end{equation}
where ${\bf \hat{T}}$ stands for an $N$ index tensor, with
components $T_{k_1...k_N}$, where $k_i=1,2,3$. The necessary and
sufficient condition for a description of the correlation function
within local realism in the general case reads
\begin{equation}
\sum_{k_1,k_2...,k_N=1}^{2} \left|
\alpha(1)_{k_1}\alpha(2)_{k_2}...\alpha(N)_{k_N}
T_{k_1k_2...k_N}\right| \leq 1. \label{27D}
\end{equation}
for any possible choice of local coordinate systems for individual
particles. Again if
\begin{equation}
\sum_{k_1,...,k_N=1}^{2}T_{k_1...k_N}^2 \leq 1 \label{28N}
\end{equation}
for any set of local coordinate systems, the $N$-qubit correlation
function can be described by a local realistic model. The proof of
these fact are generalizations of the ones presented earlier
pertaining to two particles. {The sufficient condition for
violation of the general Bell's inequality for $N$ particles by a
general state of $N$ qubits can be found in
Ref.~\cite{zukowskibrukner}.}

\subsection{Concluding remarks}
The inequalities presented above represent the full set of standard
``tight'' Bell's inequalities for an arbitrary number of parties.
Any non tight inequality is weaker than tight ones. Such Bell's
inequalities can be used to detect entanglement, not as efficiently
as entanglement general witnesses. However, they have the advantage
over the witnesses that they are systems-independent. They detect
entanglement no matter what is the actual Hilbert space that
describes the subsystems.

As we shall show below the Bell inequalities analyzed above also
show that the entanglement violating them is directly applicable in
some quantum informational protocols that beat any classical ones of
the same kind.  This will be shown via an explicit construction of
such protocols.

\section{Quantum reduction of communication complexity}
\label{secqcc}

In his review paper entitled ''Quantum Communication Complexity (A
Survey)'' Brassard~\cite{brassard} posed a question: {\it ''Can
entanglement be used to save on classical communication?''} He
continued that there are good reasons to believe at first that the
answer to the question is negative. Holevo's theorem~\cite{holevo}
states that no more than $n$ bits of classical information can be
communicated between parties by the transmission of $n$ qubits
regardless of the coding scheme as long as no entanglement is shared
between parties. If the communicating parties share prior
entanglement, twice as much classical information can be transmitted
(this is so called ''superdense coding''~\cite{bennettwiesner}), but
no more. It is thus reasonable to expect that even if the parties
share entanglement no savings in communication can be achieved
beyond that of the superdense coding ($2n$ bits per $n$ qubits
transmitted).

It is also well known that entanglement alone cannot be used for
communication.  Local operations performed on any
subsystem of an entangled composite system cannot have any
observable effect on any other subsystem; otherwise it could be
exploited to communicate faster than light. One would thus
intuitively conclude that entanglement is useless for saving
communication. Brassard, however, concluded {\it ''... all the
intuition in this paragraph is wrong.''}

The topic of classical communication complexity was introduced and
first studied by Andrew Yao in 1979~\cite{yao}. A typical
communication complexity problem can be formulated as follows. Let
Alice and Bob be two separated parties who receive some input data
of which they know only their own data and not the data of the
partner. Alice receives an input string $x$ and Bob an input string
$y$ and the goal is for both of them to determine the value of a
certain function $f(x,y)$. Before they start the protocol Alice and
Bob are even {\it allowed to share (classically correlated) random
strings} or any other data, which might improve the success of the
protocols. They are allowed to process their data locally in
whatever way. The obvious method to achieve the goal is for Alice to
communicate $x$ to Bob, which allows him to compute $f(x,y)$. Once
obtained, Bob can then communicate the value $f(x,y)$ back to Alice.
It is the topic of communication complexity to address the
questions: {\it Could there be more efficient solutions for some
functions $f(x,y)$? What are these functions?}

A trivial example that there could be more efficient solutions then
the obvious one given above is a constant function $f(x,y)\!=\!c$,
where $c$ is a constant. Obviously here Alice and Bob do not need to
communicate at all, as they can simply take $c$ for the value of the
function. However there are functions for which the only obvious
solution is optimal, that is only transmission of $x$ to Bob
warrants that he reaches the correct result. For instance, it is
shown that $n$ bits of communication are necessary and sufficient
for Bob to decide whether or not Alice's $n$-bit input is the same
as his one~\cite{brassard,kushilevitz}.

Generally one might distinguish the following two types of communication
complexity problems:
\begin{enumerate}
\item What is the minimal amount of communication
(minimal number of bits) required for the parties to determine the
value of the function with certainty?
\item What is the highest possible probability for the parties
to arrive at the correct value for the function if only a {\it
restricted} amount of communication is allowed?
\end{enumerate}
Here we will consider only the second class of problems. Note that
in this case one does not insist on the correct value of the
function to be obtained with certainty. While an error in computing
the function is allowed, the parties try to compute it correctly
with as high probability as possible.

From the perspective of the physics of quantum information
processing the natural questions is: {\it Are there communication
complexity tasks for which the parties could increase the success in
solving the problem if they share prior entanglement?} In their
original paper Cleve and Buhrman~\cite{clevebuhrman} showed that
entanglement can indeed be used to save classical communication.
They showed that to solve a certain three-party problem with
certainty the parties need to broadcast at least 4 bits of
information, in a classical protocol, whereas in the quantum
protocol (with entanglement shared) it is sufficient for them to
broadcast only 3 bits of information. This was the first example of
a communication complexity problem that could be solved with higher
success than it is be possible with any classical protocol.
Subsequently, Buhrman, Cleve and van Dam~\cite{buhrmanclevevandam}
found a two-party problem that can be solved with a probability of
success exceeding $85\%$ and 2 bits of information communicated if
prior shared entanglement is available, whereas the probability of
success in a classical protocol could not exceed $75\%$ with the
same amount of communication.

The first problem whose quantum solution requires significantly
smaller amount of communication compared to classical solutions was
discovered by Buhrman, van Dam, H{\o}yer and
Tapp~\cite{buhrmanvandamhoyertapp}. They considered a $k$-party task
which requires roughly $k \ln k$ bits of communication in a
classical protocol, and exactly $k$ bits of classical communication
if the parties are allowed to share prior entanglement. The quantum
protocol of Ref.~\cite{buhrmanclevevandam} is based on  the
violation of the CHSH inequality by two-qubit maximally entangled
state. Similarly, the quantum protocols of multi-party
problems~~\cite{buhrmanclevevandam,clevebuhrman,buhrmanvandamhoyertapp}
are based on an application of the GHZ-type argument against local
realism for multi-qubit maximally entangled states.
Galvao~\cite{galvaopra} has shown an equivalence between the CHSH
and GHZ tests for three particles and the two- and three-party
quantum protocols of Ref.~\cite{buhrmanclevevandam}, respectively.
In a series of
papers~\cite{bruknerzukowskizeilinger,bruknerzukowskipanzeilinger,expqcc,PATEREK}
it was shown that entanglement violating a Bell inequality can
always be exploited to find a better-than-any-classical solution to
some communication complexity problems. In this brief overview we
mainly follow the approach introduced in these papers. The approach
has been further developed and applied in
Ref.~\cite{augusiakhorodecki,tamir} (See also Ref.~\cite{reznik}).

\subsection{The problem and its optimal classical solution}

Imagine several spatially separated partners, $P_1$ to $P_N$, each
of whom has some data  known to him/her only, denoted here as $X_i$,
with $i=1,...,N$. They face a joint task: to compute the value of a
function $T(X_1,...,X_N)$. This function depends on all data.
Obviously they can get the value of $T$ by sending all their data to
partner $P_N$, who does the calculation and announces the result.
But are there ways to reduce the amount of communicated bits, i.e.
to reduce the communication complexity of the problem? 
%
%

Assume that every partner $P_k$ receives a two bit string
$X_k=(z_k,x_k)$ where $z_k,x_k\in \{0,1\}$. We shall consider
specific task functions which have the following form
$$T=f(x_1,...,x_N)(-1)^{\sum_{k=1}^N z_k}, $$ where $f\in \{0,1\}$ the sum in the exponent is modulo 2.
The partners know also the probability distribution (``promise'') of
the bit strings (``inputs''). There are two constraints on the
problem. Firstly, we shall consider only distributions, which are
completely random with respect to $z_k$'s, that is a class of  the
form $p(X_1,...,X_N)=2^{-N}p'(x_1,...,x_N)$. Secondly, communication
between the partners is restricted to $N-1$ bits. Assume that we ask
the last partner to give his/her answer $A(X_1,...,X_N)$, equal to
$\pm 1$, to the question what is functional value $T(X_1,...,X_N)$
in each run for the given set of inputs $X_1,...,X_N$.

For simplicity, we shall introduce now $y_k=(-1)^{z_k}$, $y_k\in
\{-1,1\}$. We shall use $y_k$ as a synonym of $z_k$. Since $T$ is
proportional to $\prod_k y_k$, the final answer $A$  is completely
random if it does not depend on {\it every} $y_k$. Thus, information
on $z_k$'s from all $N-1$ partners must somehow reach $P_N$.
Therefore the only communication ``trees'' which might lead to a
success are those in which each $P_k$ sends only a one-bit message
$m_k\in\{0,1\}$. Again we introduce: $e_k=(-1)^{m_k}$, $e_k\in
\{-1,1\}$, and will treat is as synonym of $m_k$.

The average success of a communication protocol can be measured with
the following fidelity function
\begin{equation}
F=\sum_{X_1,...,
X_N}p(X_1,...X_N)T(X_1,...X_N)A(X_1,...X_N),\end{equation}  or
equivalently
\begin{equation}
F=\frac{1}{2^N} \sum_{x_1,...,x_N=0}^{1} p'(x_1,...,x_N) f(x_1,...,
x_N) \sum_{y_1,...y_N=-1}^{1} \prod_{k=1}^N{y_k} A(x_1,...,x_N;y_1,
...,y_N ). \label{FIDELITY}
\end{equation}
The probability of success is $P =(1+F)/2$.

The first steps of a derivation of the reduced form of the fidelity
function for an optimal classical protocol will now be presented
(the reader may reconstruct the other steps or consult
references~\cite{bruknerzukowskipanzeilinger,expqcc}). In a
classical protocol the answer $A$ of the partner $P_N$ can depend on
the local input $y_N$, $x_N$, and messages, $e_{i_1},..., e_{i_l},$
received {\em directly} from a subset of $l$ partners $P_{i_1},
...,P_{i_l}$:
\begin{equation}A=A(x_N,y_N,
{e}_{i_1},..., {e}_{i_l}).\end{equation} Let us fix $x_N$, and treat
$A$ as a function $A_{x_N}$ of the remaining ${l+1}$  dichotomic
variables $$y_N,{e}_{i_1},..., {e}_{i_l}.$$ That is, we treat now
$x_N $ as a fixed index. All such functions can be thought of as
$2^{l+1}$ dimensional vectors, because the values of each such a
function form a sequence of the length equal to the number of
elements in the domain. In the $2^{l+1}$ dimensional space
containing  such functions one has {\em an orthogonal basis} given
by
\begin{equation}V_{jj_1...j_l}(y_N, {e}_{i_1},...,
{e}_{i_l})={y^j_N}\prod_{k=1}^l
{{e}^{j_k}_{i_k}},\label{QQQ}
\end{equation}
where $j,j_1,...,j_l \in \{0,1\}$. Thus, one can expand $A(x_N,y_N,
{e}_{i_1},..., {e}_{i_l})$ with respect to this basis and the
expansion coefficients read
\begin{equation}
c_{jj_1...j_l}(x_N)=\frac{1}{2^{l+1}} \sum_{y_N, {e}_{i_1},...,
{e}_{i_l}=-1}^{1} A(x_N,y_N, {e}_{i_1},..., {e}_{i_l})
{V}_{jj_1...j_l}(y_N, {e}_{i_1},..., {e}_{i_l}).\end{equation} Since
$|A|=|V_{jj_1,...j_l}|=1,$ one has $|c_{jj_1...j_l}(x_N)|\leq 1.$ We
put the expansion into the expression for $F$ and obtain
\begin{equation}
F=\frac{1}{2^N} \sum_{x_1,...,x_N=0}^{1} g(x_1,...,x_N)
\sum_{y_1,...,y_N=-1}^{1} \prod_{h=1}^N y_h
\left[\sum_{j,j_1,...j_l=0}^{1} c_{jj_1...j_l}(x_N) y_N^j
\prod_{k=1}^l {e}^{j_k}_{i_k} \right],
\end{equation}
where $g(x_1, \ldots, x_N) \equiv f(x_1, \ldots, x_N)p'(x_1, \ldots,
x_N).$ Because $\sum_{y_N=-1}^{1}y_Ny_N^0=0$, and
$\sum_{y_{k}=-1}^{1} y_{k}{e}_{k}^0=0$, only the term with all
$j,j_1,...,j_l$ equal to unity can give a non-zero contribution to
$F$. Thus, $A$ in $F$ can be replaced by
\begin{equation} A'=y_Nc_N(x_N)\prod_{k=1}^l {e}_{i_k}, \label{XXX1}\end{equation}
where $c_N(x_N)$ stands for $c_{11...1}(x_N)$. Next,  notice that,
for example, ${e}_{i_1}$, can depend only on local data $x_{i_1}$,
$y_{i_1}$ and the messages obtained by $P_{i_1}$ from a subset of
partners: $e_{p_1},..., e_{p_m}.$ This set does not contain any of
the $e_{i_k}$'s of the formula~(\ref{XXX1}) above. In analogy with
$A$, the function ${e}_{i_1}$, for a fixed $x_{i_1}$, can be treated
as a vector, and thus can be expanded in terms of orthogonal basis
functions (of a similar nature as eq.~(\ref{QQQ})), etc. Again, the
expansion coefficients satisfy $|c'_{jj_1...j_m}(x_{i_1})|\leq 1$.
If one puts this into $A'$, one obtains a
new form of $F$, which after a trivial summation over $y_N$ and
$y_{i_1}$ depends on
 $c_N(x_N)c_{i_1}(x_i)\prod_{k=2}^l {e}_{i_k},$ where
$c_{i_1}(x_i)$ stands for $c'_{11...1}(x_{i_1})$, and its modulus is
again bounded by $1$. Note that, $y_N$ and $y_{i_1}$ disappear, as
$y_k^2=1$.

As each message appears in the product only once, we continue this
procedure of expanding those messages which depend on earlier
messages, till it halts. The final reduced form of the formula for
the fidelity of an optimal protocol reads
\begin{equation}
F=\sum_{x_1,..., x_N=0}^{1} g(x_1, \ldots,
x_N)\prod_{n=1}^{N}c_n(x_n), \label{BELL}
\end{equation}
with $|c_n(x_n)|\leq 1$. Since $F$ in eq.~(\ref{BELL}) is linear in
every $c_n(x_n)$, its extrema are at the limiting values
$c_n(x_n)=\pm1$. {\em In other words, a Bell-like inequality $|F|
\leq \textrm{Max}({F})\equiv B(N)$ gives the upper fidelity bound}.
Note, that the above derivation shows that optimal classical
protocols include one in which partners $P_1$ to $P_{N-1}$ send to
$P_N$ one bit messages which encode the value of $e_k=y_kc(x_k)$,
where $k=1,2,...,N-1$.

\subsection{Quantum solutions}

The  inequality for $F$ suggests that some problems may have quantum
solutions, which surpass any classical ones in their fidelity.
Simply one may use an entangled state $|\psi\rangle$ of $N$ qubits
that violates the inequality. Send to each of the partners one of
the qubits. In a protocol run all $N$ partners make measurements on
the local qubits, the settings of which are determined by $x_k$.
They measure a certain qubit observable
$\vec{n}_k(x_k)\cdot\vec\sigma$. The measurement results
$\gamma_k=\pm1$ are multiplied by $y_k$, and the partner $P_k$, for
$1\leq k \leq N-1$, sends a bit message to $P_N$ encoding the value
of $m_k=y_k\gamma_k$. The last partner calculates
$y_N\gamma_N\prod_{k=1}^{N-1}m_k$, and announces  this as $A$. The
average fidelity of such a process is
\begin{equation}
F= \sum_{x_1, \ldots, x_N=0}^{1} g(x_1, \ldots, x_N)
\langle\psi|\otimes_{n=1}^{N}(\vec{n}_k(x_k)\cdot\vec{\sigma}_k) |\psi\rangle,\nonumber\\
\label{BELL-VIOLATED}
\end{equation}
and in certain problems can even reach {\it unity}.

For some tasks the quantum vs. classical fidelity ratio grows {\it
exponentially} with $N$. This is the case, for example, for the
so-called {\em modulo-4 sum} problem. Each partner receives a
two-bit input string $(X_k=0,1,2,3;$ $k = 1,\ldots,N).$ The promise
is that $X_k$'s are distributed such that $(\sum_{k=1}^{N}
X_k){\textrm{mod} 2} = 0.$ The task is\footnote{It can be formulated
in terms of a task function $T = 1-(\sum_{k=1}^{N} X_k)\textrm{mod}
4.$  An alternative formulation of the problem reads $f
=\cos(\frac{\pi}{2}\sum_{k=1}^{N}X_k)$ with $p'  =
\frac{1}{2^{-N+1}} |\cos(\frac{\pi}{2}\sum_{k=1}^{N}X_k)|.$}: $P_N$
must tell whether the sum modulo-4 of all inputs is  0 or 2.

For this problem  the classical fidelity bounds decrease
exponentially with $N$, that is $B(F)\leq2^{-K+1},$ where $K=N/2$
for even and $K=(N+1)/2$  for odd number of parties. If one uses the
$N$ qubit GHZ states: $|\mbox{GHZ}\rangle =
\frac{1}{\sqrt{2}}(|z+,...,z+\rangle+|z-,...,z-\rangle)$, where
$|z\pm \rangle$ is the state of spin $\pm 1$ along the $z$-axis, and
suitable pairs of local settings, the associated Bell inequality can
be violated maximally. Thus, one has a quantum protocol which always
gives the correct answer.

In all quantum protocols considered here entanglement that leads to
a violation of Bell's inequality is a resource that allows for
better-than-classical efficiency of the protocol. Surprisingly, one
can also show a version of a quantum protocol without
entanglement~\cite{galvaopra,expqcc}. The partners exchange a single
qubit, $P_k$ to $P_{k+1}$ and so on, and each of them makes a
suitable unitary transformation on it (which depends on $z_k$ and
$x_k$). The partner $P_N$, who receives the qubit as the last one,
additionally performs a dichotomic measurement. The result he/she
gets is equal to $T$. For details, including an experimental
realization see Ref.~\cite{expqcc}. The obvious conceptual advantage
of such a procedure is that the partners exchange a single qubit,
from which due to the Holevo bound~\cite{holevo} one can read out at
most one bit of information. In contrast with the protocol involving
entanglement, no classical transfer of any information is required,
except from the announcement by $P_N$ of his measurement result!

In summary, if one has a pure entangled state of many qubits (this
can be generalized to higher-dimensional systems and Bell's
inequalities involving more than two measurement settings per
observer), then there exist a Bell inequality which is violated by
this state. This inequality has some coefficients $g(x_1,...,x_n)$,
in front of correlation functions, which can always be renormalized
in such a way that
$$\sum_{x_1,...,x_n=0}^{1} |g(x_1,...,x_n)|=1.$$ The function $g$ can
always be interpreted as a product of the dichotomic function
$f(x_1,...,x_n)=\frac{g(x_1,...,x_N)}{|g(x_1,...,x_N)|} = \pm 1$ and
a probability distribution $p'(x_1,...,x_n)=|g(x_1,...,x_n)|$. Thus
we can construct a communication complexity problem that is tailored
to a given Bell's inequality, with task function $T=\prod_i^N y_i
f$. All this can be extended beyond qubits, see
Ref.~\cite{bruknerzukowskizeilinger,PATEREK}.

{As it  was shown, for three or more parties, $N \geq 3$, quantum solutions for certain communication
complexity problems can achieve probabilities of success of unity.
 This is not the case for $N=2$ and
the problem based on the CHSH inequality. The maximum quantum value for the left hand side of  the CHSH
inequality~(\ref{e:bellinequ1}) is just $\sqrt{2}-1$. This is much bigger than the Bell bound of $0$,
but still not the largest possible value, for an arbitrary theory that is not following local realism, which equals to $1$.
Because the maximum possible violation of the inequality is not attainable by
quantum mechanics several questions arise. Is this limit forced by
the theory of probability, or by physical laws? We will address this
question in the next section, and look what would  be the consequences of a
maximal  logically possible violation of the CHSH inequality.}

\subsection{Stronger-than-quantum-correlations} \label{sec:2}

The Clauser-Horne-Shimony-Holt (CHSH) inequality~\cite{chsh} for
local realistic theories gives the upper bound on a certain
combination of correlations between two space-like separated
experiments. Consider Alice and Bob who independently perform one
out of two measurements on their part of the system, such that in
total there are four experimental set-ups: $(x,y)=(0,0)$, $(0,1)$,
$(1,0)$ or $(1,1)$. For any local hidden variable theory the CHSH
inequality must hold. One can put it  the following  form:
\begin{eqnarray} \label{CHSHprob}
\lefteqn{p(a=b|x=0,y=0)+p(a=b|x=0,y=0)} \nonumber \\ & &
+p(a=b|x=0,y=0)+p(a=-b|x=0,y=0) \leq 3,
\end{eqnarray}
or equivalently,
\begin{equation}
\sum_{x,y=0,1} p(a \oplus b =x \cdot y)\leq  3. \label{nick}
\end{equation}
In the latter form we interpret the dichtomic measurement results as of binary values, $0$ or $1$, and their relations are put as  `modulo 2
sums', denoted here by $\oplus$. One has $0\oplus 0=1\oplus 1= 0$ and $0\oplus1=1$. For example,
$p(a=b|x=0,y=0)$ is the probability that Alice's and Bob's outcomes
are the same when she chooses setting $x$ and he setting $y$.

As discussed in previous sections quantum mechanical correlations
can violate the local realistic bound of inequality~(\ref{nick}) and
the limit was proven by Cirel'son~\cite{cirelson} to be
$2+\sqrt{2}$. In Ref.~\cite{popescurohrlich} Popescu and Rohrlich
asked why quantum mechanics allows a violation of the CHSH
inequality with a value of $2+\sqrt{2}$, but not more, though the
maximal logically possible value is $4$. Would a violation with a
value larger than $2+\sqrt{2}$ lead to (superluminal) signaling?.
If this were true, then quantum correlations could be understood as
maximal allowed correlations respecting non-signaling requirement.
This could give us an insight on the origin of quantum correlations,
without any use of the Hilbert space formalism.

The non-signaling condition is equivalent to the requirement that the marginals are
independent of the partner's choice of setting
\begin{eqnarray}
p(a|x,y) &\equiv& \sum_{b=0,1} p(a,b|x,y) = p(a|x),\\
p(a|x,y) &\equiv& \sum_{b=0,1} p(a,b|x,y) = p(a|x)
\end{eqnarray}
where $p(a, b|x, y)$ is the joint probability for outcomes $a$ and
$b$ to occur given $x$ and $y$ are the choices of measurement
settings, respectively and $p(a|x)$ is the probability for outcome
$a$ given $x$ is the choice of measurement setting. Popescu and
Rohrlich constructed a toy-theory where the correlations reach the
maximal algebraic value of $4$ for left hand expression of the inequality~(\ref{CHSHprob}), but
are nevertheless not in contradiction with signaling. The
probabilities in the toy model are given by
\begin{eqnarray} \label{eq:sstrong}
\left.{\begin{array}{l}
p(a=0,b=0|x,y) = \frac{1}{2} \\
p(a=1,b=1|x,y) = \frac{1}{2}
\end{array}}\right\}
& &\textrm{if~}xy\in\{00,01,10\}, \nonumber \\
& & \nonumber \\
\left.{\begin{array}{l}
p(a=1,b=0|x,y)= \frac{1}{2} \\
p(a=0,b=1|x,y) = \frac{1}{2}
\end{array}}\right\} & &
 \textrm{if~}xy=11.
\end{eqnarray}
Indeed one has
\begin{equation} \label{sstrong}
\sum_{x,y=0,1} p(a \oplus b = x\cdot y)=4.
\end{equation}

Van Dam~\cite{vandam}  and independently Cleve considered how
plausible are stronger-than-quantum correlations from the point of
view of communication complexity, which describes how much
communication is needed to evaluate a function with distributed
inputs. It was shown that the existence of correlations that
maximally violate the CHSH inequality would allow to perform all
distributed computations (between two parties)  of dichotomic
functions with a communication constraint to just one bit. If one is
ready to believe that nature should not allow for ``easy life''
concerning communication problems, this could be a reason why
superstrong correlations are indeed not possible.

Instead of superstrong correlations one usually speaks about a
``nonlocal box'' (NLB) or Popescu-Rohrlich (PR) box, as an imaginary
device that takes as inputs $x$ at Alice's and $y$ at Bob's side,
and outputs $a$ and $b$ at respective sides, such that $a \otimes b
= x\cdot y$. Quantum mechanical measurements on a maximally
entangled state allow for a success probability of
$p=\cos^2\frac{\pi}{8}=\frac{2+\sqrt{2}}{4} \approx 0.854$ at the
game of simulating NLBs. Recently, it was shown that in any
``world'' in which it is possible to implement an approximation to
the NLB, that works correctly with probability greater than
$\frac{3+\sqrt{6}}{6}=90.8\%$,  for all distributed computations of
dichotomic functions with a one-bit  communication constraint, one
can find a protocol that gives always the correct values,
Ref.~\cite{brassardcomm}. This bound is an improvement over van
Dam's one, but still has a gap with respect to the bound imposed by
quantum mechanics.

\subsubsection{Superstrong correlations trivializes communication
complexity}

We shall present  a proof that availability of a perfect NLB would allow for a
solution of a general communication complexity problem for a binary function, with an
exchange of a single bit of information. The proof is due to van
Dam~\cite{vandam}.

Consider a Boolean function $f : \{0,1\}^n \times \{0,1\}$, which
has as inputs two $n$-bit strings $\vec{x} = (x_1,...,x_n)$ and
$\vec{y} = (y_1,...,y_n)$. Suppose that Alice receives the $\vec{x}$
string and Bob, who is separated from Alice, the $\vec{y}$-string,
and they are to determine the function value $f(\vec{x},\vec{y})$ by
communicating as little as possible. They have, however, NLBs as
resources.

First, let us notice that any dichotomic function
$f(\vec{x},\vec{y})$ can be rewritten as a finite summation:
\begin{equation}
f(\vec{x},\vec{y}) = \sum_{i=1}^{2^n} P_i(\vec{x}) Q_i(\vec{y}),
\end{equation}
where $P(\vec{x})$ are polynomials in $\vec{x}\in \{0,1\}$ and
$Q_i(\vec{y}) = y^{i_1}_1\cdot ... \cdot y^{i_n}_n$ are monomials in
$y_i \in \{0,1\}$ with $i_1,...,i_n \in \{0,1\}$. {Note that the latter ones
constitute an orthogonal basis in a $2^n$ dimensional space.
The decomposed function $f$ is treated as a function of $y$'s, while the inputs $x_1,...,x_n$
are considered as indices numbering functions $f$.}
 Note that there are
$2^n$ different monomials. Alice can locally compute all the $P_i$
values by herself and likewise Bob can compute all $Q_i$ by himself.
These values determine the settings of Alice and Bob that will be
chosen in $i$-th run of the experiment. Note that to this end  they need
in general exponentially many NLBs. Alice and Bob perform for every
$i\in \{1,...,2^n\}$ a measurement on the $i$-th NLB in order to
obtain without any communication a collection of bit values $a_i$
and $b_i$, with the property $a_i \otimes b_i = P_i(\vec{x})
Q_i(\vec{y}) $. Bob can add all his $b_i$ to $\sum_{i=1}^{2^n} b_i$
values without requiring any information from Alice, and he can
broadcast this single bit to Alice. She, on her part, computes the
sum of her $a_i$ to $\sum_{i=1}^{2^n} a_i$ and adds Bob's bit to it.
The final result
\begin{equation}
\sum_{i=1}^{2^n} (a_i \oplus b_i) =  \sum_{i=1}^{2^n} P_i(\vec{x})
Q_i(\vec{y}) = f(\vec{x},\vec{y})
\end{equation}
is the function value. Thus, {\it superstrong correlations
trivialize every communication complexity problem.}

\section{The Kochen-Specker Theorem}
\label{sec:6}

{ In previous sections we have seen, that tests of Bell's
inequalities are not only theory independent tests of
non-classicality, but also have applications in quantum information
protocols. Examples are communication complexity
problems~\cite{bruknerzukowskipanzeilinger}, entanglement
detection~\cite{hyllus}, security of key distribution~\cite{acin},
and quantum state discrimination~\cite{schmidt}. Thus entanglement
which violates local realism can be seen as a resource for efficient
information processing. Can quantum contextuality -- the fact that
quantum predictions disagree from the ones of non-contextual
hidden-variable theories -- also be seen as such a resource? We will
give an affirmative answer to this question by considering explicit
examples of a quantum game.}

The Kochen-Specker theorem is a "no go" theorem that proves a
contradiction between predictions of quantum theory and those of
{\it non-contextual} hidden variable theories. It was proved by Bell
in 1966~\cite{bell1966} and independently by Kochen and Specker in
1967~\cite{kochen-specker}. The non-contextual hidden-variable
theories are based on the conjecture of the following three
assumptions:
\begin{enumerate}
\item {\it Realism}: It is a
model that allows one to use all variables $A_{m}(n)$ in the
theoretical description of the experiment, where $A_{m}(n)$ gives
the value of some observable $A_m$ which {\em could} be obtained if
the knob setting were at positions $m$. The index $n$ describes the
entire experimental ``context'' in which $A_m$ is measured and is
operationally defined through the positions of all other knob
settings in the experiment, which are used to measure other
observables jointly with $A_m$. All $A_{m}(n)$'s are treated as
perhaps unknown, but still fixed, (real) numbers, or variables for
which a proper joint probability distribution can be defined.

\item {\it Non-contextuality}: The value assigned to an observable $A_m(n)$ of an individual system is
independent of the experimental context $n$ in which it is measured,
in particular of any properties that are measured {\it jointly} with
that property. This implies that $A_{m}(n) = A_m$ for all contexts
$n$.
\item ``Free will''. The experimenter is free to choose the observable and its context. The choices are independent of the actual hidden values of $A_m$'s, etc.
\end{enumerate}
Note that ``non-contextuality'' implies locality (i.e.,
non-contextuaily with respect to a remote context), but there is no
implication other way round. One might have theories which are
local, but locally non-contextual.

It should be stressed that the local realistic and non-contextual
theories provide us with predictions which can be tested
experimentally, and which can be derived {\it without making any
reference to quantum mechanics} (though many derivations in the
literature give exactly the opposite impression). In order to
achieve this, it is important to realize that predictions for
noncontextual realistic theories can be derived in a completely
operational way~\cite{simon}. For concreteness, imagine that an
observer wants to perform a measurement of an observable, say the
 square, $S^2_{\vec{n}}$, of a spin component of a spin-1 particle along a certain
direction $\vec{n}$. There will be an experimental procedure for
trying to do this as accurately as possible. We will refer to this
procedure by saying that one sets the ``control switch`` of his/her
apparatus to the position $\vec{n}$. In all experiments that we will
discuss only a finite number of different switch positions is
required. By definition different switch positions are clearly
distinguishable for the observer, and the switch position is all he
knows about. Therefore, in an operational sense the measured
physical observable is entirely defined by the switch position. From
the above definition it is clear that the same switch position can
be chosen again and again in the course of an experiment. Notice
that in such an approach as described above, it does not matter which
observable is ``really'' measured and to what precision. One just
derives general predictions, provided that certain switch positions
are chosen.

In the original Kochen-Specker proof~\cite{kochen-specker}, the
observables that are considered are squares of components of the
spin 1 along various directions. Such observables have values 1 or
0, as the components themselves have values 1,0, or $-1$. The squares
of spin components $\hat{S}^2_{\vec{n}_1}$, $\hat{S}^2_{\vec{n}_2}$
and $\hat{S}^2_{\vec{n}_3}$ along any three orthogonal directions
$\vec{n}_1$, $\vec{n}_2$, and $\vec{n}_3$ can be measured jointly.
Simply, the corresponding quantum operators commute with each other.
In the framework of a hidden-variable theory one assigns to an
individual system a set of numerical values, say $+1$,0,$+1$,... for the
square of spin component along each direction $S^2_{\vec{n}_1}$,
$S^2_{\vec{n}_2}$, $S^2_{\vec{n}_3}$,... that can be measured on the
system. If any of the observables is chosen to be measured on the
individual system, the result of the measurement would be the
corresponding value. In a non-contextual hidden variable theory one
has to assign to an observable, say $S^2_{\vec{n}_1}$, the {\it
same} value independently of whether it is measured in an
experimental procedure jointly as a part of some set
$\{S^2_{\vec{n}_1}, S^2_{\vec{n}_2}, S^2_{\vec{n}_3}\}$ or of some
other set $\{S^2_{\vec{n}_1}, S^2_{\vec{n}_4}, S^2_{\vec{n}_5}\}$ of
physical observables, where  $\{\vec{n}_1, \vec{n}_2, \vec{n}_3\}$
and $\{\vec{n}_1, \vec{n}_4, \vec{n}_5\}$ are triads of orthogonal
directions. Notice that within quantum theory some of the operators
corresponding to the observables from the first set may {\it not
commute} with some corresponding to the observables from the second
set.

The squares of spin components along orthogonal directions satisfy
\begin{equation}
\hat{S}^2_{\vec{n}_1}+ \hat{S}^2_{\vec{n}_2} + \hat{S}^2_{\vec{n}_3}
= s(s+1) =2. \label{spin1}
\end{equation}
This is {\em always} so for a particle of spin 1 (s=1). This implies
that for every measurement of three squares of mutually orthogonal
spin components two of the results will be equal to one, and one of
them will be equal to zero. The Kochen-Specker theorem considers a
set of triads of orthogonal directions $\{\vec{n}_1, \vec{n}_2,
\vec{n}_3\}$, $\{\vec{n}_1, \vec{n}_4, \vec{n}_5\}$,..., for which
at least some of the directions have to appear in several of the
triads. The statement of the theorem is that there are sets of
directions for which it is not possible to give any assignment of
1's and 0's to the directions consistent with the
constraint~(\ref{spin1}). The original theorem
in~\cite{kochen-specker} used 117 vectors, but this has subsequently
been reduced to 33 vectors~\cite{peres1991} and 18
vectors~\cite{cabello}. Mathematically the contradiction with
quantum predictions has its origin in the fact that the classical
structure of non-contextual hidden variable theories is represented
by commutative algebra, whereas quantum mechanical observables need
not be commutative, making it impossible to embed the algebra of
these observables in a commutative algebra.

The disproof of noncontextually relies on the assumption that the
same value is assigned to a given physical observable,
$\hat{S}^2_{\vec{n}}$, regardless with which two other observables
the experimenter chooses to measure it. In quantum theory the
additional observables from one of those sets correspond to
operators that do not commute with the operators corresponding to
additional observables from the other set.  As it was stressed in a
masterly review on hidden variable theories by Mermin~\cite{mermin},
Bell wrote~\cite{bell1966} that ``These different possibilities
require different experimental arrangements; there is no {\it a
priori} reason to believe that the results ... should be the same.
The result of observation may reasonably depend not only on the
state of the system (including hidden variables) but also on the
complete disposition apparatus.'' Nevertheless, as Bell himself
showed, the disagreement between predictions of quantum mechanics
and of the hidden-variables theories can be strengthened if
non-contextuality is replaced by a much more compelling assumption
of locality. Note that in Bohr's doctrine of the inseparability of
the object and the measuring instrument, an observable {\it is}
defined through the entire measurement procedure applied to measure
it. Within this doctrine one would not speak about measuring the
same observable in different contexts, but rather about measuring
entirely different maximal observables, and deriving from it the
value of a degenerate observable. Note that Kochen-Specker argument
necessarily involves degenerate observables. This is why it does not
apply to single qubits.

\subsection{A Kochen-Specker Game}

We will now consider a quantum game which is based on the
Kochen-Specker argument strengthened by the locality condition (See
Ref.~\cite{svozil}). We consider a pair of entangled spin 1
particles, which form a singlet state with total spin 0. A formal
description of this state is given by
\begin{equation}
|\Psi\rangle = \frac{1}{\sqrt{3}} (|1\rangle_{\vec{n}}
|-1\rangle_{\vec{n}} + |-1\rangle_{\vec{n}} |1\rangle_{\vec{n}} -
|0\rangle_{\vec{n}} |0\rangle_{\vec{n}}), \label{statespin0}
\end{equation}
where, for example, $|1\rangle_{\vec{n}} |-1\rangle_{\vec{n}}$ is
the state of the two particles with spin projection +1 for the first
particle and spin projection -1 for the second particle 1 along the
same direction $\vec{n}$. It is important to note that this state is
invariant under a change of the direction $\vec{n}$. { This
implies that if the spin components for the two particles are measured
along an arbitrary direction, however the same both sides, the sum of the two local results is always
zero. This is a direct consequence of the conservation of
angular momentum.}

We now present the quantum game introduced in Ref.~\cite{cleve}. The
requirement in the proof of the Kochen-Specker theorem can be
formulated as the following problem in geometry. There exists an
explicit set of vectors $\{\vec{n}_1, . . . , \vec{n}_m\}$ in
$\mathbf{R}^3$ that cannot be colored in red (i.e., assign the value
1 to the spin squared component along that direction) or blue (i.e.,
assign the value 0) such that both of the following conditions hold:
\begin{enumerate}
\item For every orthogonal pair of vectors $\vec{n}_1$ and $\vec{n}_2$, they are not
both colored red.
\item For every mutually orthogonal triple of vectors $\vec{n}_i$, $\vec{n}_j$, and $\vec{n}_k$, at least one of them is colored
red.
\end{enumerate}
For example, the set of vectors can consist of 117 vectors from the
original Kochen-Specker proof~\cite{kochen-specker}, 33 vectors from
Peres's proof or 18 vectors from Cabello's proof~\cite{cabello}.

The Kochen-Specker game employs the above sets of
vectors. Consider two separated parties, Alice and Bob. Alice
receives a random triple of orthogonal vectors as her input and Bob
receives a single vector randomly chosen from the triple as his
input. Alice is asked to give a trit indicating which of her
three vectors is assigned color 1 (implicitly, the other two vectors
are assigned color 0). Bob outputs a bit assigning a color to his
vector. The requirement is that Alice and Bob assign the same color
to the vector that they receive in common.  Nevertheless, it is straightforward to
show that the existence of a perfect classical strategy in which
Alice and Bob can share classically correlated strings for this game
would violate the reasoning used in the Kochen-Specker theorem.  On the other hand, there is a perfect quantum
strategy using the entangled state (\ref{statespin0}). If Alice and Bob
share two particles in this state, Alice can perform a measurement of
squared spin components pertaining to directions
$\{\vec{n}_i,\vec{n}_j,\vec{n}_k\}$, which are equal to those of the three input vectors, and Bob measures
squared spin component in direction $\vec{n}_l$ for his input. Then
Bob's measurement will necessarily yield the same answer as the
measurement by Alice along the same direction.

Concluding this section we note that quantum contextuality is also
closely related to quantum error correction~\cite{divincenzo},
quantum key distribution~\cite{nagata}, one-location quantum
games~\cite{aharon}, and entanglement detection between internal
degrees of freedom.

\subsection{Temporal Bell's Inequalities (Leggett-Garg Inequalities)}

In the last section we will consider one more basic information
processing task, random access code problem. It can be solved with a
quantum set-up with a higher efficiency than it is classically
possible. We will show that the resource for better-than-classical
efficiency is a violation of ``temporal Bell's inequalities''  --
the inequalities that are satisfied by temporal correlations of
certain class of hidden-variable theories. Instead of considering
correlations between measurement results on distantly located
physical systems, here we focus on one and the same physical system
and analyze correlations between measurement outcomes at different
times. The inequalities were first introduced by Leggett and
Garg~\cite{leggettgarg} in the context of testing superspositions of
macroscopically distinct quantum states. Since our aim here is
different, we will look at general assumptions that allows us to
derive temporal Bell's inequalities irrespectively of whether the
object under consideration is macroscopic or not. This is why our
assumptions differ from the original ones of
Ref.~\cite{leggettgarg}. Compare also
Ref.~\cite{paz,shafiee,bruknervedral}

We consider the theories which are based on the conjunction of the
following four assumptions\footnote{There is one more difference
between the present approach and this of Ref.~\cite{leggettgarg}.
While there the observer measures a single observable having a
choice between different times of measurement, here at any given
time the observer has a choice between two (or more) different
measurement settings. One can use both approaches to derive temporal
Bell' inequalities.}:
\begin{enumerate}
\item {\it Realism}: It is a
model that allows one to use all variables $A_{m}(t)$ $m=1,2,...$ in
the theoretical description of the experiment performed at time $t$,
where $A_{m}(t)$ gives the value of some observable which {\em
could} be obtained if it were measured at time $t$. All $A_{m}(t)$'s
are treated as perhaps unknown, but still fixed numbers, or
variables for which a proper joint probability distribution can be
defined.
\item {\it Non-invasiveness}: The value assigned to an observable $A_m(t_1)$ at time $t_1$ is
independent whether or not a measurement was performed at some
earlier time $t_0$ or which observable $A_n(t_0)$ $n=1,2,...$ at
that time was measured. In other words, (actual or potential)
measurement values $A_m(t_1)$ at time $t_1$ are {\it independent} of
the measurement settings chosen at earlier times $t_0$.
\item {\it Induction}: The standard arrow of time is assumed. In
particular, the values $A_{m}(t_0)$ at earlier times $t_0$ do not
depend on the choices of measurement settings at later times
$t_1$\footnote{Note that this already follows from the
```non-invasiveness'' when applied symmetrically to both arrows of
time.}.
\item {\it ``Free will''}: The experimenter is free to choose the observable. The choices
are independent of the actual hidden values of $A$'s, etc.
\end{enumerate}

Consider an observer and allow her to choose at time $t_0$ and at
some later time $t_1$ to measure one of two dichotomic observables
$A_1(t_i)$ and $A_2(t_i)$, $i \in \{0,1\}$. The assumptions given
above imply existence of numbers for $A_1(t_i)$ and $A_2(t_i)$, each
taking values either +1 or -1, which describe the (potential or
actual) predetermined result of the measurement. For the temporal
correlations in an individual experimental run the following
identity holds: $A_1(t_0) [ A_1(t_1)-A_2(t_1) ] + A_2(t_0) [
A_1(t_1)+A_2(t_1) ] = \pm 2$. With similar steps as in derivation of
the standard Bell's inequalities, one easily obtains:
\begin{equation}
p(A_0A_0=1)+p(A_0A_1=-1) +p(A_1A_0=1)+p(A_1A_1=1) \leq 3,
\label{tchsh}
\end{equation}
where we omit the dependence on time.

An important difference between quantum contextuality and temporal
Bell's inequalities is that later can also be tested on single
qubits or {\it two-dimensional} quantum systems. We will now
calculate the temporal correlation function for consecutive
measurements of a single qubit. Take an arbitrary mixed state of a
qubit, written as $ \rho = \frac{1}{2} (\mathbf{1} + \vec{r}\cdot
{\vec \sigma})$, where $\mathbf{1}$ is the identity operator, ${\hat
\sigma}\equiv (\vec{\sigma_x},\vec{\sigma_y},{\vec\sigma_z})$ are
the Pauli operators for three orthogonal directions $x$, $y$ and
$z$, and $\vec{r}\equiv (r_x,r_y,r_z)$ is the Bloch vector with the
components $r_i\!=\!\mbox{Tr}(\rho \vec{\sigma_i})$.

Suppose that the measurement of the observable $\vec{\sigma} \cdot
\vec{a}$ is performed at time $t_0$, followed by the measurement of
$\vec{\sigma} \cdot \vec{b}$ at $t_1$, where $\vec{a}$ and $\vec{b}$
are directions at which spin is measured. The quantum correlation
function is given by $ E_{QM}(\vec{a},\vec{b}) = \sum_{k,l=\pm 1} k
\cdot l \cdot \mbox{Tr}(\rho \pi_{\vec{a},k}) \cdot
\mbox{Tr}(\pi_{\vec{a},k} \pi_{\vec{b},l}),$ where, e.g.,
$\pi_{\vec{a},k}$ is the projector onto the subspace corresponding
to the eigenvalue $k\!=\!\pm 1$ of the spin along $\vec{a}$. Here we
use the fact that after the first measurement the state is projected
on the new state $\pi_{\vec{a},k}$. Therefore, the probability to
obtain the result $k$ in the first measurement and $l$ in the second
one is given by $\mbox{Tr}(\rho \pi_{\vec{a},k})
\mbox{Tr}(\pi_{\vec{a},k} \pi_{\vec{b},l})$. Using $\pi_{\vec{a},k}=
\frac{1}{2} ( \mathbf{1} + k \vec{\sigma} \cdot \vec{a})$ and
$\frac{1}{2} \mbox{Tr} [(\vec{\sigma} \cdot \vec{a}) (\vec{\sigma}
\cdot \vec{b})]= \vec{a} \cdot \vec{b}$ one can easily show that the
quantum correlation function can simply be written as
\begin{equation}
E_{QM}(\vec{a},\vec{b}) = \vec{a} \cdot \vec{b}.\label{eqm}
\end{equation}
Note that in contrast to the usual correlation function the temporal
one (\ref{eqm}) does not dependent of the initial state $\rho$. Note
also that a slight modification of our derivation of Eq. (\ref{eqm})
can also apply to the cases in which the system evolves between the
two measurements following an arbitrary unitary transformation.

The scalar product form of quantum correlations~(\ref{eqm}) allows
for the violation of the temporal Bell inequality and the maximal
value of the left-hand side of~(\ref{tchsh}) is achieved for the
choice of the measurement settings: $\vec{a}_1 \!=\!
\frac{1}{\sqrt{2}} (\vec{b}_1 - \vec{b}_2)$, $\vec{a}_2 \!=\!
\frac{1}{\sqrt{2}} (\vec{b}_1 + \vec{b}_2) $ and is equal to
$2+\sqrt{2}$.

\subsection{Quantum Random Access Codes}

Random access code is a communication task for two parties, whom we
call again Alice and Bob. Alice receives some classical $n$-bit
string known only to her (her local input). She is allowed to send
just a one bit message, $m$, to Bob. Bob is asked to tell the value
of the $b$-th bit of Alice, $b=1,2...,n$. However $b$ is known only
to him (this is his local input data). The goal is to construct a
protocol enabling Bob  to tell the value $b$-th  bit of Alice, with
as high average probability of success as possible, for a uniformly
random distribution of Alice's bit-strings, and a uniform
distribution of $b$'s. Note that, since Alice does not know in
advance which bit Bob is to recover. Thus she has no option to send
just this required bit.

If they share a quantum channel then one speaks about a quantum
version of the previous problem. Alice is asked to encode her
classical $n$-bit message into 1 qubit (quantum bit) and send it to
Bob. He performs some measurement on the received qubit to extract
the required bit. In general, the measurement that he uses will
depend on which bit he wants to reveal. The idea behind these
so-called quantum random access codes already appeared in a paper
written circa 1970 and published in 1983 by Stephen Wiesner
\cite{wiesner}.

We illustrate the concept of random access code with the simplest
scheme, in which in a classical framework Alice needs to encode a
two-bit string $b_0b_1$ into a single bit, or into a single qubit in
a quantum framework.

In the classical case Alice and Bob need to decide on a protocol
defining which bit-valued message is to be sent by Alice, for each
of the four possible values of her two-bit string $b_0b_1$. There
are only $2^4 = 16$ different deterministic protocols, thus the probability
of success can be evaluated in a straightforward way. The optimal
deterministic classical protocols can then be shown to have a
probability of success $P_C = 3/4.$ For example, if Alice sends one
of the two bits, then Bob will reveal this bit with certainty and
have probability of $1/2$ to reveal the other one. Since any
probabilistic protocol can be represented as a convex combination of
the 16 deterministic protocols, the corresponding probability of
success for any such probabilistic protocol will be given by the
weighted sum of the probabilities of success of the individual
deterministic protocols. This implies that the optimal probabilistic
protocols can at best be as efficient as the optimal deterministic
protocol, which is $3/4$.

Ambainis {\it et al.}~\cite{ambainis} showed that there is a quantum
solution of the random access code with probability of success $P_Q
= \cos^2(\pi/8)\approx 0.85$. It is realized as follows: depending
on her two-bit string $b_0b_1$, Alice prepares one of the four
states $|\psi_{b_0b_1}\rangle$. These states are chosen to be on the
equator of the Bloch sphere, separated by equal angles of $\pi/2$
radians (see figure 3). Using the Bloch sphere parametrization
$|\psi(\theta,\phi)\rangle = \cos(\theta/2)|0\rangle +
\exp{(i\phi)}\sin(\theta/2)|1\rangle$, the four encoding states are
represented as:
\begin{eqnarray}
|\psi_{00}\rangle &=& |\psi(\pi/2,\pi/4)\rangle, \nonumber \\
|\psi_{01}\rangle &=& |\psi(\pi/2,7\pi/4)\rangle, \nonumber \\
|\psi_{10}\rangle &=& |\psi(\pi/2,3\pi/4)\rangle, \nonumber \\
|\psi_{11}\rangle &=& |\psi(\pi/2,5\pi/4)\rangle.
\end{eqnarray}
Bob's  measurements, which he uses to guess the bits, will depend on which bit he wants to
obtain. To guess $b_0$, he projects the qubit along the $x$-axis in
the Bloch sphere, and to decode $b_1$ he projects it along the
$y$-axis. He then estimates the bit value to be $0$ if the
measurement outcome was along the positive direction of the axis and
$1$ if it was along the negative axis. It can easily be calculated
that the probability of successful retrieving of the correct bit
value is the same in all cases: $P_Q = \cos^2(\pi/8)\approx 0.85$,
which is higher than the optimal probability of success $P_C = 0.75$
of the classical random access code using one bit of communication.

\begin{figure}[t]
\sidecaption[t]
\includegraphics[scale=.35]{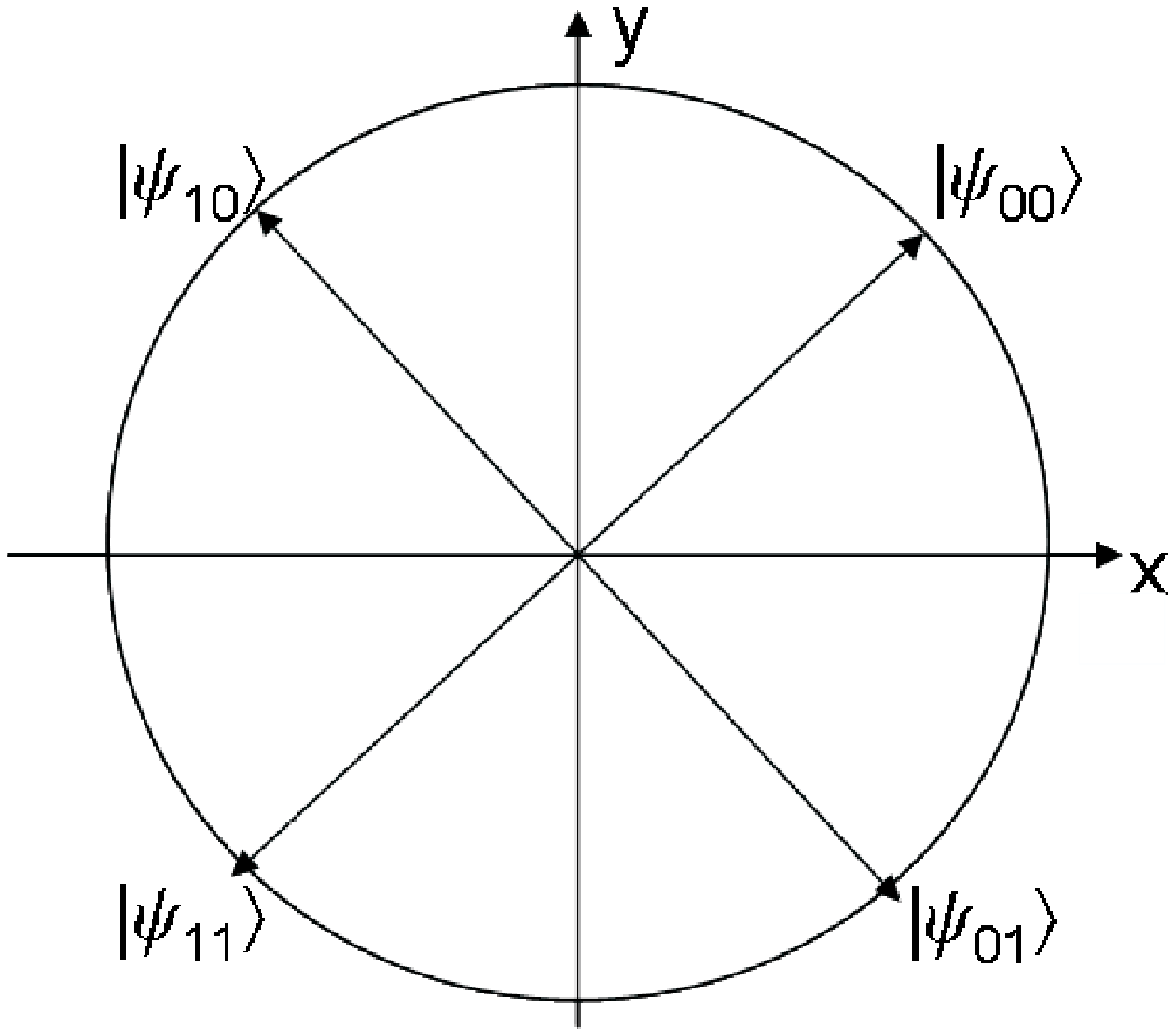}
\caption{The set of encoding states and decoding measurements in
quantum random access code represented in the $x-y$ plane of the Bloch sphere.
Alice prepares one of the four quantum states $\psi_{b_0b_1}$ to
encode two bits $b_0, b_1 \in \{0,1\}$. Depending on which bit Bob
wants to reveal he performs either a measurement along the $x$ (to
reveal $b_0$) or along the $y$ axis (to reveal $b_1$).}
\label{fig:3}       
\end{figure}

We will now introduce a hidden variable model of the quantum
solution to see that the key resource in its efficiency lies in
violation of temporal Bell's inequalities. Galvao~\cite{galvaodiss}
was the first to point to the relation between violation of Bell's type
inequalities and quantum random access codes. See also
Ref.~\cite{spekkens} for a relation with the parity-oblivious
multiplexing.

A hidden-variable model equivalent to the quantum protocol, which
best fits the temporal Bell's inequalities can be put as a
description of the following modification of the original quantum
protocol. Alice prepares the initial state of her qubit as a
completely random state, described by a density matrix proportional
to the unit operator, $\sigma_0$. Her parity of bit values  $b_0
\oplus b_1$ defines a measurement basis, which is used by her to
prepare the
state to be sent to Bob. 
Note that the result of the dichotomic measurement in the basis
defined by $b_0\oplus b_1$ is, due to the nature of the initial
state, completely random, and totally uncontrollable by Alice. To
fix the bit value $b_1$ (and thus also the value $b_0$, since the
parity is defined by the choice of the measurement basis) on her
wish, Alice either leaves the state unchanged, if the result of
measurement corresponds to her wish of $b_1$ or she rotates the
state in the $x-y$ plane at $180^{o}$ to obtain the orthogonal
state, if the result corresponds to $b_1 \oplus 1$. Just a glance at
the states involved in the standard quantum protocol shows what are
the two complementary (unbiased) bases which define her measurement
settings, and which resulting states are linked with which values of
$b_0b_1$. After the measurement the resulting state is sent to Bob,
while Alice is in possession of a bit pair $b_0b_1$, which is
perfectly correlated with the qubit state on the way to Bob. That
is, we have exactly the same starting point as in the original
quantum protocol.

Now, it is obvious that the quantum protocol violates the temporal
inequalities, while any hidden variable model of the above
procedure, using the four assumptions (1.-4.) behind the temporal
inequalities is not violating them. What is important the saturation
of the temporal inequalities is equivalent to a probability of
success of $3/4$.

The link with temporal Bell's inequalities points onto another
advantage of quantum over classical random access codes. Usually,
one considers the advantage to be only {\it resource dependent}.
With this we mean that there is an advantage as far as one compares
one classical bit with one qubit. Yet, the proof given above shows
that quantum strategy has an advantage over {\it all} hidden
variable models respecting (1.-4.), i.e. also those where Alice and
Bob use systems with arbitrarily large number of degrees of freedom.

Concluding this section and the Chapter we would like to point onto
an interesting research avenue. Here we gave a brief review on the
results demonstrating that ``no go theorems'' for various hidden
variable classes of theories, are behind better-than-classical
efficiency in many quantum communication protocols. It would be
interesting to investigate the link between fundamental features of
quantum mechanics and the power of quantum computation. It has been
shown that temporal Bell's inequalities distinguish between
classical and quantum search (Grover) algorithm~\cite{morikoshi}.
Cluster states -- a resource for measurement-based {\it quantum}
computation (also known as ``one-way'' quantum computation) in which
information is processed by a sequence of adaptive single-qubit
measurements on the state -- are shown to violate Bell's
inequalities~\cite{scaranietal,ggb}. Similarly, the CSHS and GHZ
problems are shown to be closely related to measurement-based {\it
classical} computation, as does the Popescu-Rohrlich
box~\cite{anders}. These results point on the aforementioned link
but we are still far away from understanding what are the key
non-classical ingredients that give rise to the enhanced quantum
computational power. The question gets even more fascinating after
realizing that not only too
low~\cite{anders,nest,brenner,vidal,markov,fannes} but also too much
entanglement does not allow  powerful quantum
computation~\cite{eisertetal,winter}.

\begin{acknowledgement}
We acknowledge support from the Austrian Science Foundation FWF
within Project No. P19570-N16, SFB and CoQuS No. W1210-N16 and the
European Commission, Project QAP (No. 015848). The collaboration is
a part of an \"{O}AD/MNiSW program.

\end{acknowledgement}

\input{references}

\end{document}

%% file: references.tex
%
%
%